\documentclass[11pt]{article}
\usepackage{graphicx} 
\begin{document}

\title{The 1/1 resonance in Extrasolar Systems: Migration from planetary to satellite orbits}

\author{John D. Hadjidemetriou and  George Voyatzis \\ Derpartment of Physics, University of Thessaloniki \\
              Thessaloniki, Greece \\
              email: hadjidem@auth.gr, voyatzis@auth.gr }        
\maketitle

\begin{abstract}
We present families of symmetric and asymmetric periodic orbits at the 1/1 resonance, for a planetary system consisting of a star and two small bodies, in comparison to the star, moving in the same plane under their mutual gravitational attraction. The stable 1/1 resonant periodic orbits belong to a family which has a planetary branch, with the two planets moving in nearly Keplerian orbits with non zero eccentricities and a satellite branch, where the gravitational interaction between the two planets dominates the attraction from the star and the two planets form a close binary which revolves around the star. The stability regions around periodic orbits along the family are studied. Next, we study the dynamical evolution in time of a planetary system with two planets which is initially trapped in a stable 1/1 resonant periodic motion, when a drag force is included in the system. We prove that if we start with a 1/1 resonant planetary system with large eccentricities, the system migrates, due to the drag force, {\it along the family of periodic orbits} and is finally trapped in a satellite orbit. This, in principle, provides a mechanism for the generation of a satellite system: we start with a planetary system and the final stage is a system where the two small bodies form a close binary whose center of mass revolves around the star.

{\bf keywords :} 1/1 resonance, periodic orbits, co-orbital motion, planetary migration
\end{abstract}

\section{Introduction}

In the present work we study the dynamical evolution of an extrasolar planetary system which is close to the 1/1 resonance. It is known (Hadjidemetriou 2002; Hadjidemetriou 2006; Beaug\'e et al. 2003) that families of periodic orbits exist in the three body problem, in a rotating frame, consisting of a star (with a large mass) and two bodies with small masses (planets or satellites) that interact gravitationally. We consider here the planar case, i.e. all bodies move in the same plane (see Fig. \ref{rotating}a). In Hadjidemetriou et al. (2009), families of stable and unstable periodic orbits are found for planetary systems at the 1/1 mean motion resonance in the rotating frame . These families determine critically the topology of the phase space and consequently affect the evolution of the system. We remark that these families of 1/1 resonant periodic orbits studied here do not emanate from Trojan like orbits (e.g. Schwarz et al. 2009) and are called by some authors ``quasi-satellite orbits'' (Mikkola et al. 2006; Giuppone et al. 2010).

In this work we consider that, in addition to the gravitational forces between the three bodies, non conservative forces also act on the planets. If we assume that the planetary system is not yet fully developed and a proto-planetary nebula exists, then the motion of the planets is affected by the drag which is due to the interaction between the planets and the proto-planetary nebula. Such a dissipation effect lasts until this nebula is dissolved and the system takes its final form. This kind of study has been made in order to explain the large eccentricities, or the very close proximity of the planets, and the resonance trapping in several observed extrasolar planetary systems  (Beaug\'e and Ferraz-Mello 1993; Gomes 1996; Ferraz-Mello et al. 2003; Nelson and Papaloizou 2003a; Nelson and Papaloizou 2003b; Papaloizou, 2003; Beaug\'e et al. 2006; Morbidelli et al. 2007; Zhou et al. 2008; Hadjidemetriou and Voyatzis 2010). 

A planetary system under non conservative forces evolves, in the phase space, following the {\it stable part} of the families of periodic orbits, as has been found for the 2/1 and 3/1 resonances by Ferraz-Mello et al. (2003); Beaug\'e et al. (2006); Hadjidemetriou and Voyatzis (2010, 2011). In the present case of the 1/1 resonance we start with a {\it planetary orbit} of the two small bodies, with large eccentricities and follow its evolution under non conservative forces. It is shown that the system evolves along the stable family and finally can be trapped in a {\it satellite orbit}.

In section 2 we introduce our model and discuss the main dynamical issues of the conservative system, namely the planar three body problem in a rotating frame. In section 3 we present the periodic orbits of the conservative system for various ratios of the planetary masses and we study the stability regions around the stable symmetric family. In section 4 we present the results of our numerical simulations of the dissipative system and illustrate the evolution. Finally, we conclude in section 5.

%%%%%%%%%%%%%%%%%%%%%%%%%%%%%%%%%%%%%%%%%%%%%%%%%%%%%%%%%%%%%
%%% SECTION 2
%%%%%%%%%%%%%%%%%%%%%%%%%%%%%%%%%%%%%%%%%%%%%%%%%%%%%%%%%%%%%

\section{The conservative and the dissipative models}

\subsection{The model in the inertial frame of reference}

We consider the star $S$, with mass $m_0$ and the two planets $P_1$ and $P_2$, with masses $m_1$ and $m_2$, respectively, moving in the same plane, in an inertial frame where the center of mass of the system is fixed at the origin of a coordinate system $X\Omega Y$. In the conservative model we have four degrees of freedom, since the position of the system is determined by the coordinates $X_1,Y_1$ and $X_2,Y_2$ of the two planets (the position of the star is obtained from the fact that the center of mass is at $\Omega$). 

It is assumed that the nebula, which introduces the drag, rotates differentially with Keplerian circular velocity (at each radius $r$) and the non conservative force is a linear drag law (a Stokes like force) proportional to the relative velocity of the planets with respect to the nebula:
\begin{equation}
\vec R =- 10^{-n}(\vec v-\vec v_c),
\label{dis-law}
\end{equation}
where $\vec v$ is the velocity of the planet and $\vec v_c$ is the circular velocity of the nebula, given by 
\begin{equation}
\vec v_c = \sqrt{\frac{\displaystyle Gm_0}{\displaystyle r}}\: \vec e_\theta.
\label{circ-vel}
\end{equation}
The component $-10^{-n}\vec v$ in Eq. \ref{dis-law} is the purely dissipative force and the component $-10^{-n}(-\vec v_c)$ is a forcing force, due to the rotation of the nebula, which is imposed to the system. This means that the nonconservative force given by Eq. \ref{dis-law} may be either positive or negative, depending on the relative velocity of the planet with respect to the circular velocity at that point. This type of drag force has been used by Beaug\'e et al. 2006, to study migration at the 2/1 resonance. In the following we shall call the force given by Eq. \ref{dis-law} {\it dissipative}, with the meaning that the dissipation may be either positive or negative, as explained above.

The differential equations of the motion of the planets are 
\begin{equation}
\begin{array}{l}
 \ddot X_1=-m_0\frac{\displaystyle X_1-X_0}{\displaystyle r_{01}^3}-m_2\frac{\displaystyle X_1-X_2}{\displaystyle r_{12}^3}+\frac{\displaystyle R_{1x}}{\displaystyle m_1}, \\[0.6cm] 
 
  \ddot Y_1=-m_0\frac{\displaystyle Y_1-Y_0}{\displaystyle r_{01}^3}-m_2\frac{\displaystyle Y_1-Y_2}{\displaystyle r_{12}^3}+\frac{\displaystyle R_{1y}}{\displaystyle m_1}, \\[0.6cm] 
  
   \ddot X_2=-m_0\frac{\displaystyle X_2-X_0}{\displaystyle r_{02}^3}-m_1\frac{\displaystyle X_2-X_1}{\displaystyle r_{12}^3}+\frac{\displaystyle R_{2x}}{\displaystyle m_2} , \\[0.6cm] 
   
 \ddot Y_2=-m_0\frac{\displaystyle Y_2-Y_0}{\displaystyle r_{02}^3}-m_1\frac{\displaystyle Y_2-Y_1}{\displaystyle r_{12}^3}+\frac{\displaystyle R_{2y}}{\displaystyle m_2} , 
  \end{array}
\label{eq-inert}
\end{equation}
and include the gravitational interaction between the bodies and also the dissipative force acting on each planet. $r_{01}$, $r_{02}$ and $r_{12}$ are the distances between $S$ and $P_1$, $S$ and $P_2$ and $P_1$ and $P_2$, respectively. ($R_{1x}$, $R_{1y}$) and ($R_{2x}$, $R_{2y}$) are the components of the dissipative force given in Eq. \ref{dis-law}, acting on the planets $P_1$ and $P_2$, respectively. In the numerical integration of the system we assume that the center of mass of the system is fixed, because the mass of the star is much larger than the masses of the planets and the dissipative force acting on the planets is very small.

 The unit of mass is the total mass, $m$, of the system and the gravitational constant $G$, is also taken equal to unity:
$$
m=m_0+m_1+m_2=1,\ G=1.
%\label{G}
$$
\subsection{The conservative model in the rotating frame} \label{SectionConsModel}

\begin{figure}
\begin{center}
$\begin{array}{cccc}
\includegraphics[width=3.8cm,height=3.3cm]{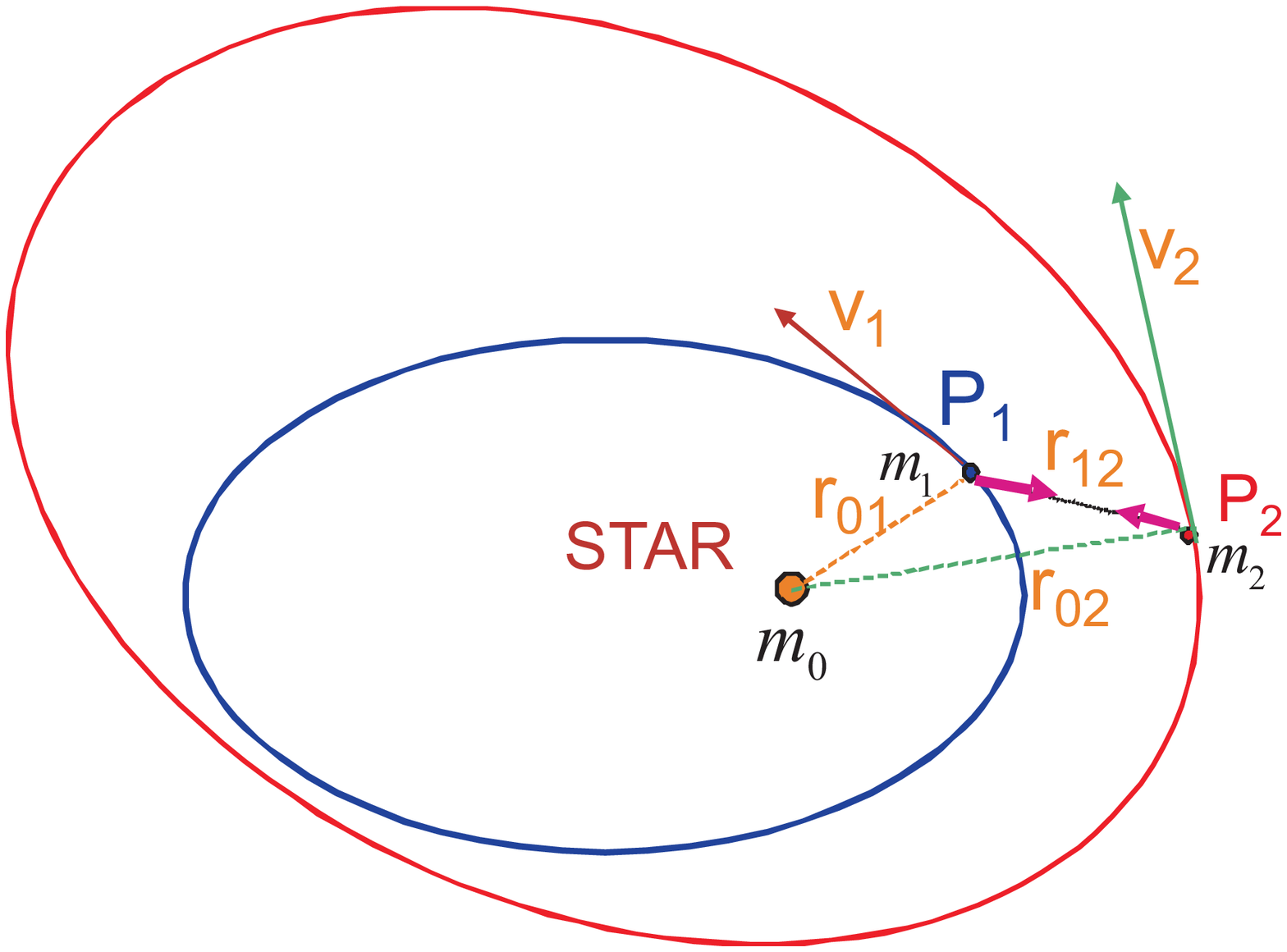} & 
\includegraphics[width=3.8cm,height=3.3cm]{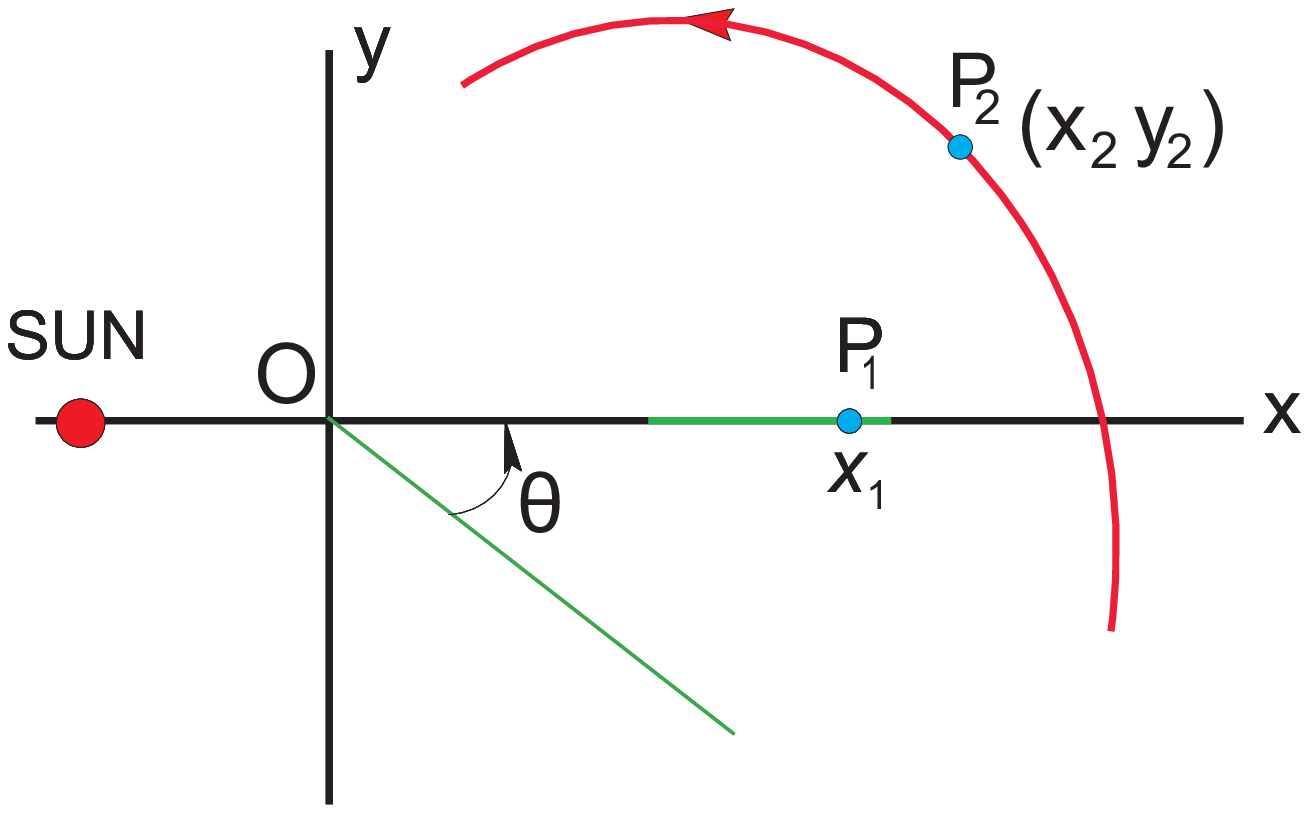}&
\includegraphics[width=3.8cm,height=3.3cm]{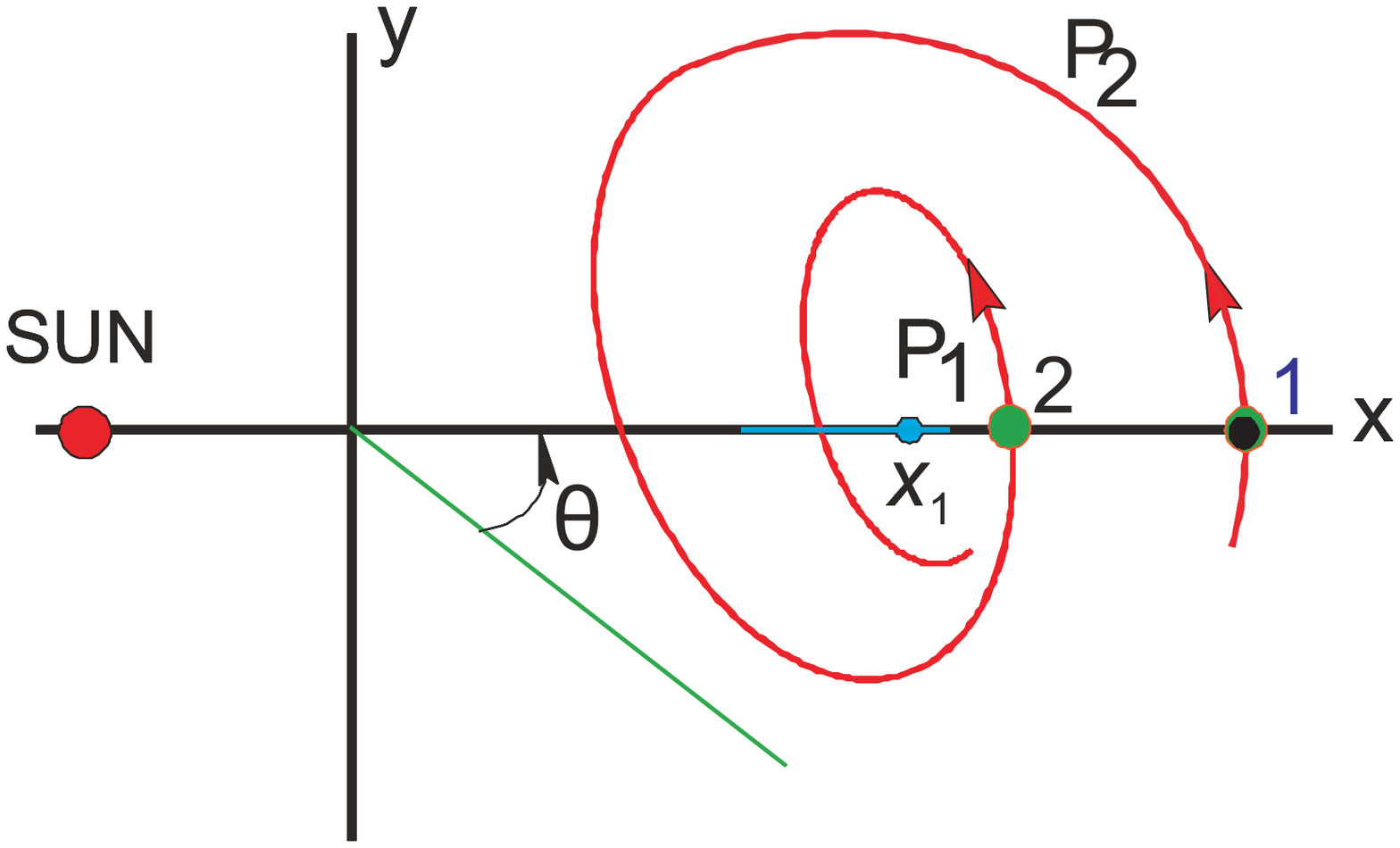} &\\
\textnormal{(a)} &\textnormal{(b)} &\textnormal{(c)}
\end{array} $
\end{center}
\caption { {\bf a} The two planets in elliptic orbits that interact gravitationally. {\bf b} The rotating frame $xOy$. The origin $O$ is at the center of mass of $S$ and $P_1$ and the $x$-axis is the line $S-P_1$. The planet $P_2$ moves in this rotating frame. The angle $\theta$ is ignorable. {\bf c} The Poincar\'e map on the surface of section $y_2=0$.}
\label{rotating} 
\end{figure}
We introduce now a {\it rotating frame} $xOy$, whose origin is the center of mass of the star $S$ and the planet $P_1$ and the $x$ axis is the line $S-P_1$ (see Fig. \ref{rotating}b). In this rotating frame the planet $P_1$ moves always on the $x$ axis and the planet $P_2$ moves in the $xOy$ plane. We still have four degrees of freedom, with variables ($x_1, x_2, y_2, \theta$), where $\theta$ is the angle between the $x$ axis and a fixed direction in inertial frame and defines the orientation of the rotating frame. The Lagrangian of the {\it conservative part} of the system, in the rotating frame, is (Hadjidemetriou 1975)
%%%
%%%
\begin{equation}
L = \frac{\displaystyle 1}{\displaystyle 2} (m_0+m_1)\left\{
q(\dot x_1^2+x_1^2\dot \theta^2)+\frac{\displaystyle m_2}{\displaystyle m}\left[\dot x_2^2+\dot y_2^2+\dot \theta^2(x_2^2+y_2^2)+2\dot\theta(x_2\dot y_2-\dot x_2y_2)\right] \right\}-V,
\label{lagrangian}
\end{equation}
where
\begin{equation}
V=-\frac{\displaystyle Gm_0 m_1}{\displaystyle r_{01}}-\frac{\displaystyle Gm_0 m_2}{\displaystyle r_{02}}-\frac{\displaystyle Gm_1 m_2}{\displaystyle r_{12}}, \nonumber
%\label{potential}
\end{equation}
and $q=m_1/m_0$.

The angle $\theta$ is ignorable and consequently the angular momentum $p_\theta=\partial{L}/\partial{\dot\theta}$, given by
\begin{equation}
p_\theta=(m_0+m_1)\left\{\dot \theta\left[qx_1^2+\frac{\displaystyle m_2}{\displaystyle m}(x_2^2+y_2^2)\right ]+\frac{\displaystyle m_2}{\displaystyle m}(x_2\dot y_2-\dot x_2y_2) \right\},
\label{angmom}
\end{equation}
is constant. We can use now the angular momentum integral to reduce the number of degrees of freedom from four to three, by eliminating the ignorable angle $\theta$. The new Lagrangian is the Routhian function (see Pars 1965)
\begin{equation}
R=\frac{\displaystyle 1}{\displaystyle 2}\left\{
q\dot x_1^2+\frac{\displaystyle m_2}{\displaystyle m}(\dot x_2^2+\dot y_2^2)-
\frac{\displaystyle 
\left [\frac{\displaystyle p_\theta}{\displaystyle (m_1+m_2)}-
\frac{\displaystyle m_2}{\displaystyle m}(x_2\dot y_2-\dot x_2y_2)\right]^2}{\displaystyle q x_1^2+\frac{\displaystyle m_2}{\displaystyle m}(x_2^2+y_2^2)}
\right\}-V.
\label{ruthian}
\end{equation}
In this way we restrict our study in the rotating frame only, in the variables ($x_1, x_2, y_2$) and the six dimensional phase space ($x_1, x_2, y_2, \dot x_1, \dot x_2, \dot y_2$). Note that $p_\theta$ appears as a fixed parameter in the Routhian (\ref{ruthian}).

In the numerical study of the evolution of the system we use the full system (Eqs. \ref{eq-inert}) and transform the motion in the rotating frame $xOy$. All the computations were performed by the Bulirch - Stoer integration method, with an accuracy of $10^{-14}$.

In order to avoid unnecessary details in the computations and restrict the study to the general features only, we use the Poincar\'e map on the surface of section
\begin{equation}
y_2=0, \;\; \dot y_2>0.
\label{map}
\end{equation}
By the Poincar\'e map (Fig. \ref{rotating}c) we reduce by one the dimensions of the phase space, which is now the five dimensional space ($x_1, x_2, \dot x_1, \dot x_2, \dot y_2$). In the following sections we present the results of the computations in projections in different coordinate planes or, equivalently, in projections in the orbital elements plane, mainly the eccentricity plane.

\subsection{Periodic orbits}

As mentioned above, we can restrict the study of the motion of a planetary system in the rotating frame of Fig. \ref{rotating}b and it is known that periodic orbits exist in this rotating frame, which belong to one dimensional families (Hadjidemetriou 2006). The initial conditions of an orbit in the rotating frame are
\[
\mathbf{X}(0)=\{x_{10}, x_{20}, y_{20}, \dot x_{10},\dot x_{20}, \dot y_{20}\},
\]
and a periodic orbit of period $T$ is defined by the condition $\mathbf{X}(0)=\mathbf{X}(T)$. Due to the system's symmetry $\Sigma=(t\rightarrow -t, x\rightarrow x, y\rightarrow -y)$, if the initial conditions $\mathbf{X}(0)$ correspond to a periodic orbit then the initial conditions 
\[
\mathbf{X}'(0)=\{x_{10}, x_{20}, -y_{20}, -\dot x_{10}, -\dot x_{20}, \dot y_{20}\},
\]
correspond also to a new ``mirror image'' periodic orbit. If the periodic orbit coincides with its mirror image orbit then it is ``symmetric'', otherwise it is called ``asymmetric'' (H\'enon, 1997; Voyatzis and Hadjidemetriou, 2005).     

In particular, for symmetric periodic orbits it is $y_{20}=0$, $\dot x_{10}=0$  $\dot x_{20}=0$, i.e. the planet $P_2$ starts from the $x$-axis perpendicularly at $t=0$ and at the same time the planet $P_1$ (that moves on the $x$-axis) is at rest. Consequently, the nonzero initial conditions of a symmetric periodic orbit at $t=0$ are
\begin{equation}
\{x_{10}, \ \ x_{20},\ \   \dot y_{20}\}.
\label{initcond}
\end{equation}

In a symmetric periodic orbit it is  at $t=0$, $\omega_i=0$ or $\pi$ and $M_i=0$ or $\pi$ ($i=1,2$) and consequently $\Delta\omega=\omega_2-\omega_1$ and $\Delta M=M_2-M_1$ are always equal to $0$ or $\pi$. Note that the configurations ($M_1=0^\circ$, $M_2=\pi$) and ($M_1=\pi$, $M_2=0^\circ$) are equivalent, separated by half a period $T$, due to the 1/1 resonance. However, along a family of asymmetric periodic orbits $\Delta\omega$ and $\Delta M$ vary.  In the following we shall present the evolution of a planetary system in its phase space, by giving the projection in coordinate planes, or in the orbital elements space and mainly in the eccentricity space.

%%%%%%%%%%%%%%%%%%%%%%%%%%%%%%%%%%%%%%%%%%%%%%%%%%%%%%%%%%%%%
%%% SECTION 3
%%%%%%%%%%%%%%%%%%%%%%%%%%%%%%%%%%%%%%%%%%%%%%%%%%%%%%%%%%%%%

\section{The dynamics of the conservative model}
In this section we consider the conservative model described in section \ref{SectionConsModel}. First we present the different families of periodic orbits of the system and then we study the stability regions in phase space close to the periodic orbits. 

\subsection{Symmetric families of 1/1 resonant periodic orbits} \label{SectionFams}
\begin{figure}
\begin{center}
\includegraphics[width=6cm, height=6cm]{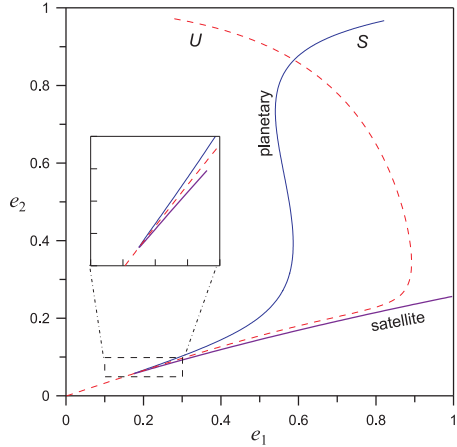} 
\end{center}
\caption{The stable ($S$) and the unstable ($U$) family of 1/1 resonant periodic orbits for the mass ratio $\rho=3$  ($m_1=0.0005$, $m_2=0.0015$). The $S$ family shows a cusp that separates the family in two parts, the planetary and satellite part.}
\label{FigFams3}
\end{figure}

Hadjidemetriou et al. (2009) have found two families of symmetric 1/1 resonant periodic orbits, one stable and one unstable. On the stable family the periastra of the two planets are in opposite directions ($\Delta \omega = \pi$) and the planets at $t=0$ are in periastron and apoastron, respectively ($\Delta M =\pi$). On the unstable family the periastra are in the same direction ($\Delta \omega = 0$) and the planets are also in periastron and apoastron, respectively ($\Delta M =\pi$). A typical example of these families, for the masses $m_1=0.0005$, $m_2=0.0015$ is presented in Fig. \ref{FigFams3} in the projection space of the eccentricities $e_1-e_2$. 

%%%%%%%%%%%%%%%%%%%%%%%%%%%%%%%%

The periodic orbits of the stable family have been called ``quasi-satellite orbits'' (Giuppone et al. 2010).  This type of orbits are also present in the restricted three body problem, beside the tadpole and horseshoe Trojan orbits (see Mikkola et al. 2006 and references therein). In Fig. \ref{FigFams3} it is clear that the stable family has a cusp as it approaches the lower eccentricities. This cusp, which is not apparent in the semi-analytical model used by Giuppone et al. (2010), divides the family in two parts, before and after the cusp, respectively. In the first part the two planets do not come close to each other, due to the resonance protection, and their mutual gravitational interaction is relatively weak. Thus, the orbits are almost Keplerian and we call them orbits of ``planetary type''. After the cusp, the planets are very close to each other and their mutual gravitational interaction dominates, forming a close binary which revolves around the star. We call these orbits of ``satellite type''. The distinction between planetary and satellite orbits will become clear in section 4, Figs. \ref{orbits123} and \ref{orbits34}.

\begin{figure}
\begin{center}
$\begin{array}{ccc}
\includegraphics[width=5.5cm]{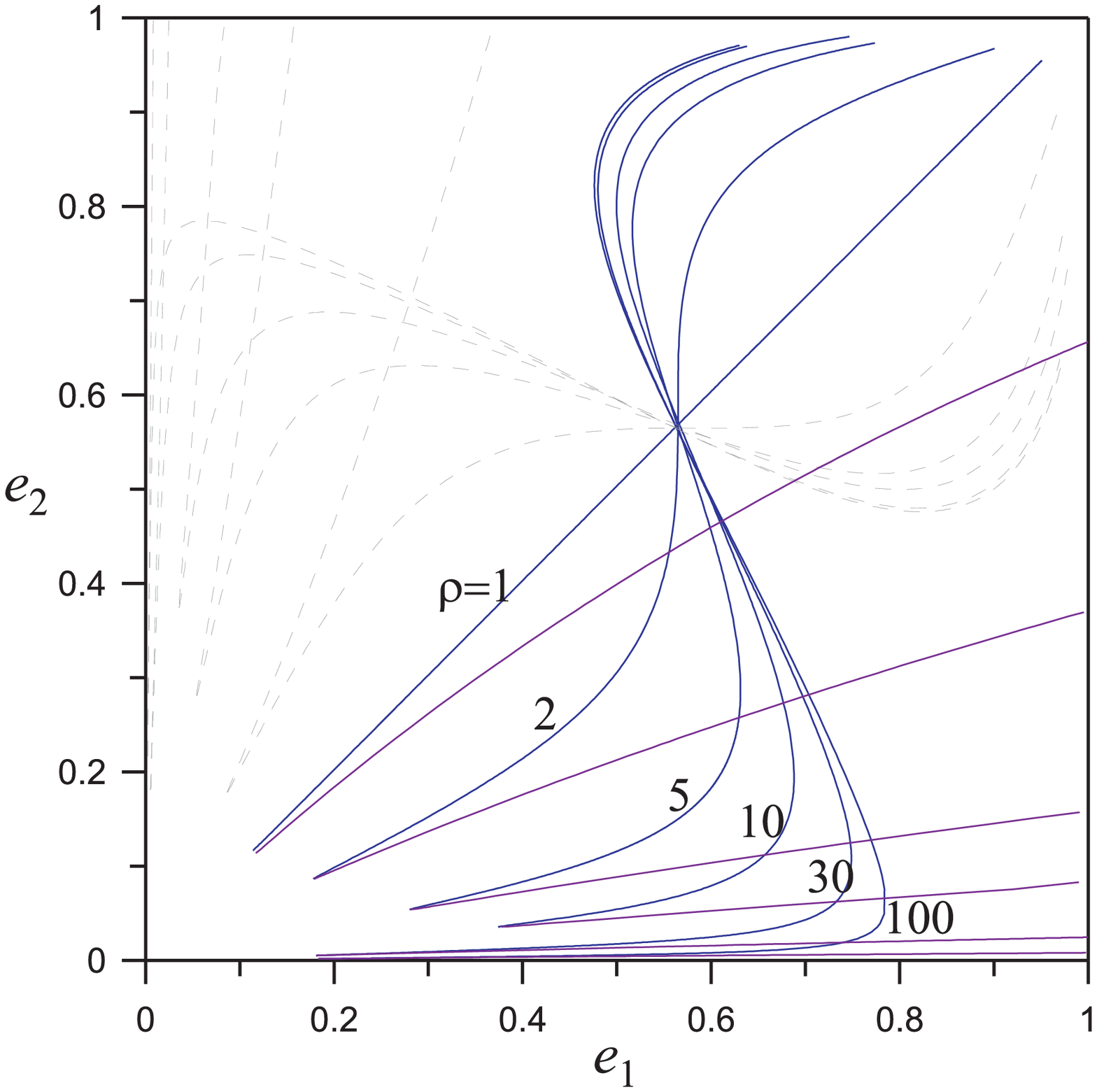} & \qquad &
\includegraphics[width=5.5cm]{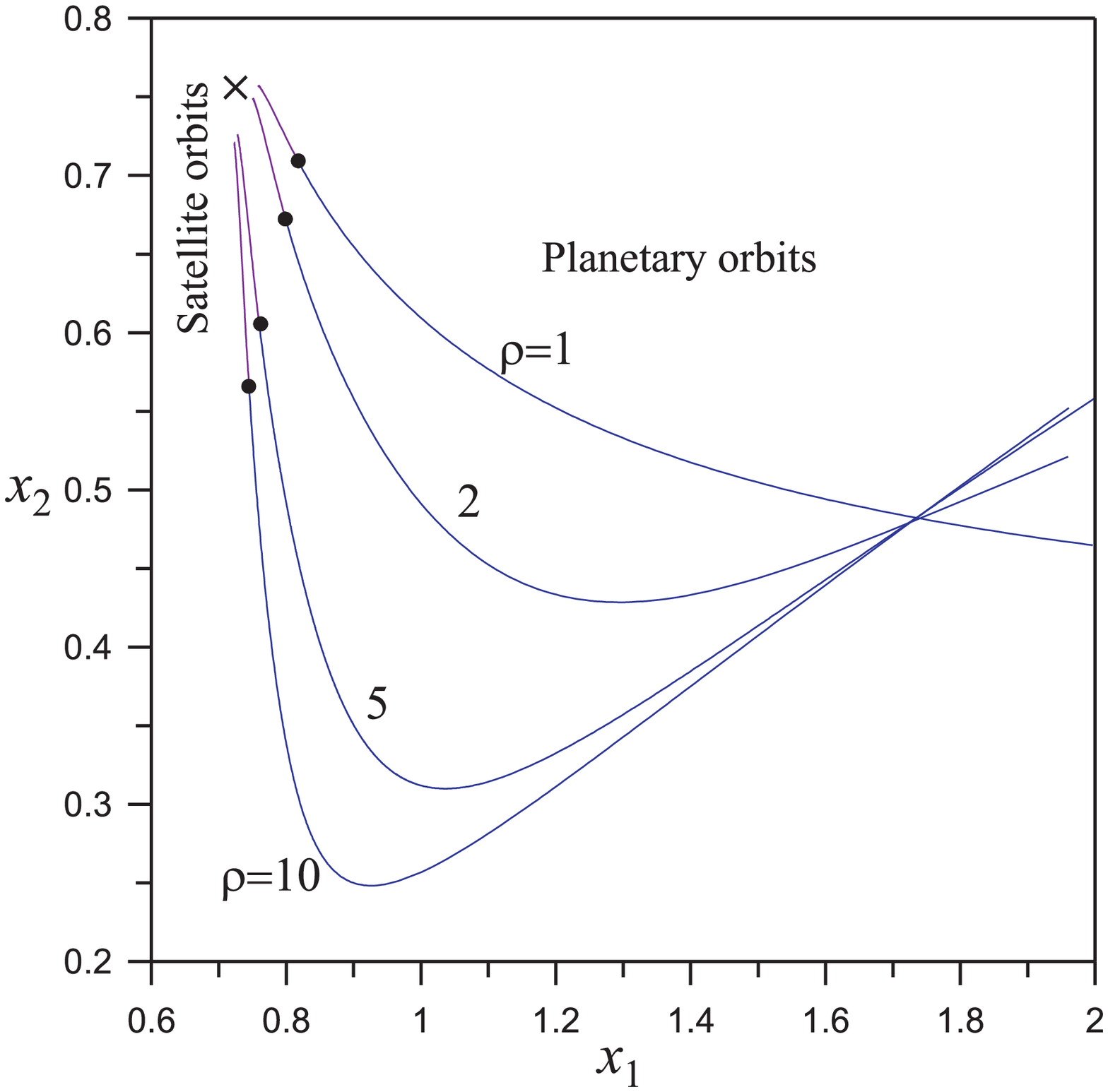}\\
\textnormal{(a)} & \qquad & \textnormal{(b)}
\end{array} $
\end{center}
\caption{{\bf a} The stable symmetric families in the eccentricity plane for various mass ratios $\rho$. In all cases the planetary and satellite parts are well distinct. The gray dashed curves indicate the symmetrical families for mass ratio $1/\rho$. {\bf b} The stable families in the plane of rotating variables $x_1-x_2$. The corresponding characteristic curves are smooth. The bold dots indicate the orbits where the cusps appear in the $e_1-e_2$ presentation. The cross indicates a planetary collision point.}
\label{FigFamsS}
\end{figure}

For small planetary masses, the location of the families of periodic orbits in the $e_1-e_2$ plane is affected only by the planetary mass ratio $\rho=m_1/m_2$ and not the total planetary mass $m_1+m_2$ (provided that $m_i \ll m_0$).   In Fig. \ref{FigFamsS}a we present the families of periodic orbits for various values of $\rho$ in the eccentricity plane and for $m_1=0.001$. Due to the intrinsic symmetry of the 1:1 resonance, the family curves that correspond to the mass ratio $\rho$ are symmetrical to the curves of mass ratio $1/\rho$ with respect to the axis $e_1=e_2$. For any value of $\rho$ the cusp is apparent dividing the family in the ``planetary'' and ``satellite'' parts. We note that the location of the planetary parts of the families are in a very good agreement with those computed by Giuppone et al. (2010) and the property that all families pass very close to the point $e_1\approx e_2\approx 0.57$ is verified. 

The cusp formed along a stable family in the eccentricity plane appears because the mutual gravitation of the planets becomes very large and the elliptic shape of the orbits is significantly deformed. Thus the osculating eccentricities are not appropriate variables to describe smoothly the family of such non Keplerian orbits. We use the osculating elements for the satellite part just to show the sharp transition from planetary to satellite motion. This cusp does not exist if we present the family in the space of initial conditions  (\ref{initcond}), expressed in the coordinates of the rotating frame $xOy$. This implies that the family is a unique family. Indeed, in Fig. \ref{FigFamsS}b the stable families are presented in the plane $x_1-x_2$ by smooth curves which seem to terminate at a planetary collision.     

\begin{figure}
\begin{center}
$\begin{array}{ccc}
\includegraphics[width=5.5cm,height=5.5cm]{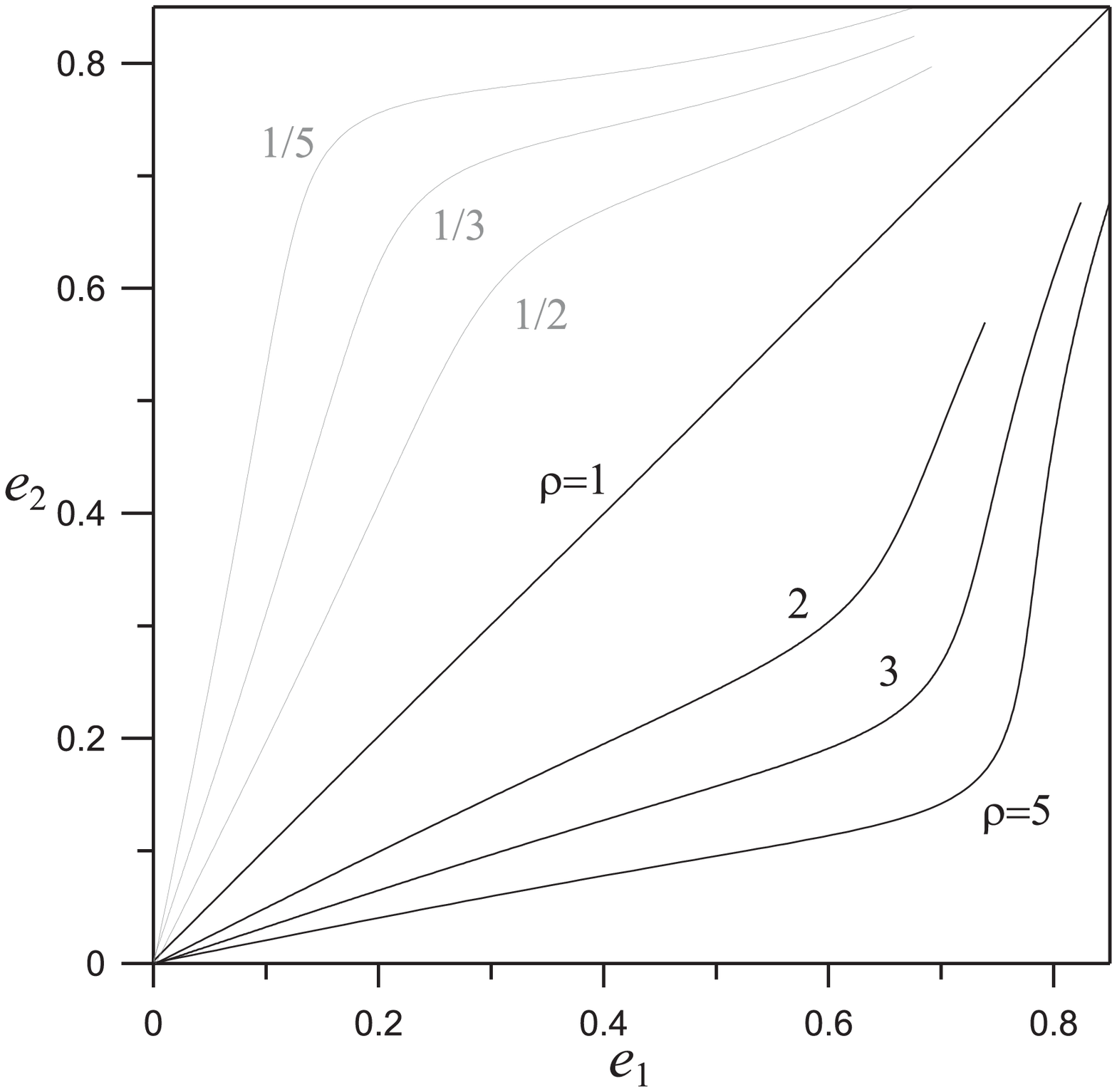} &  \qquad &
\includegraphics[width=5.5cm,height=5.5cm]{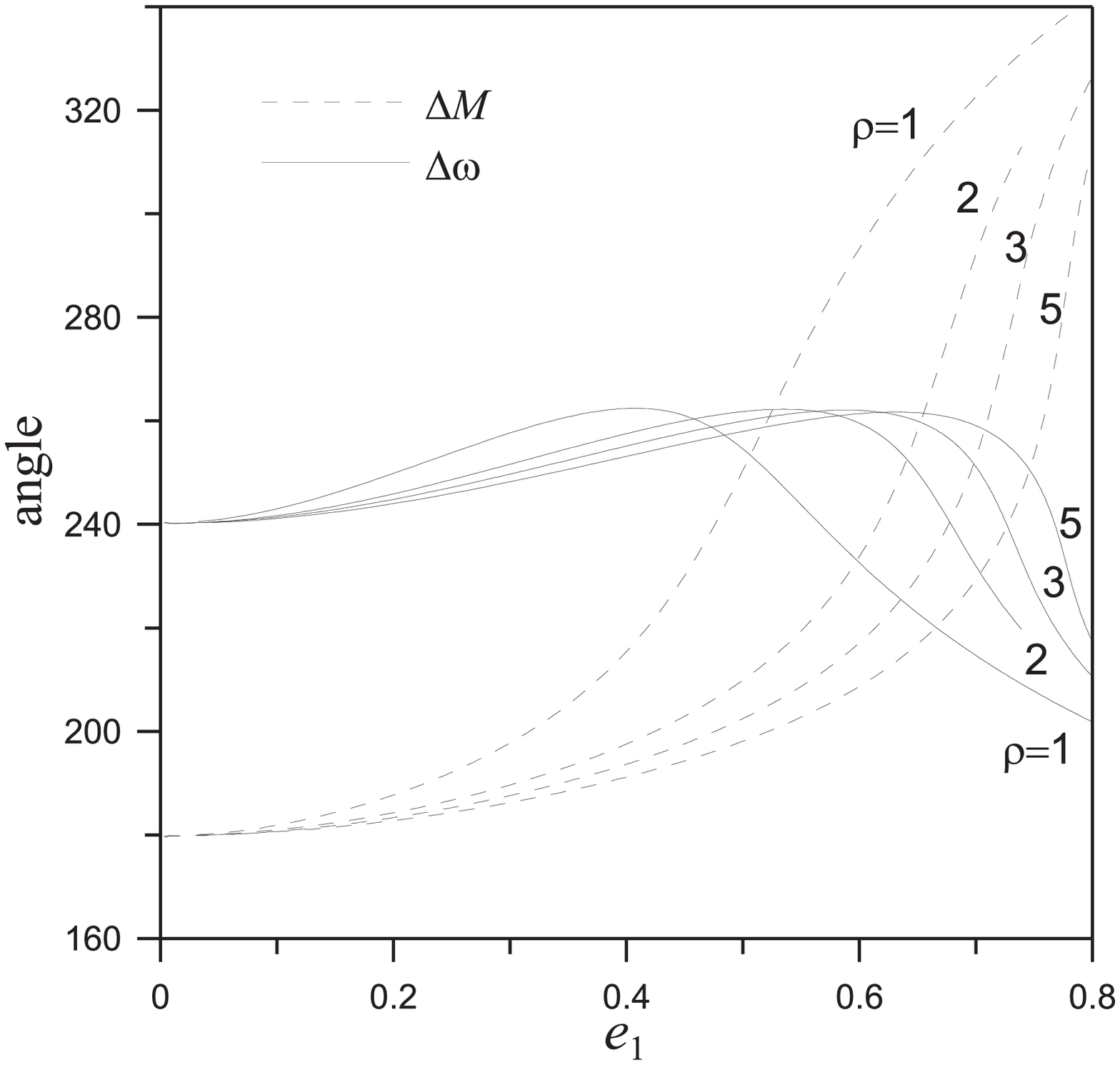}\\
\textnormal{(a)} &  \qquad & \textnormal{(b)}
\end{array} $
\end{center}
\caption{{\bf a} Asymmetric families in the eccentricity plane and for the indicated mass ratio values $\rho$. {\bf b} The variation of the angles $\Delta\omega=\omega_2-\omega_1$ and $\Delta M=M_2-M_1$ along the asymmetric families.}
\label{FigFamsA}
\end{figure}

\subsection{Asymmetric families of 1/1 resonant periodic orbits} \label{SectionFamA}
In Giuppone et al. (2010), two families ($AL_4$ and $AL_5$) of stable asymmetric periodic orbits are found by using a semi-analytical averaged model. These asymmetric orbits encompass the Lagrangian points $L_4$ and $L_5$. In this study we present such orbits in the rotating frame (Fig. \ref{rotating}b) of the general three body model. The family $AL_5$ is the mirror image of the family $AL_4$ and thus we present only one family. In Fig. \ref{FigFamsA}a the asymmetric families for some values of the mass ratios $\rho$ and $m_1=0.001$ are shown in the $e_1-e_2$ plane. All families start from zero eccentricities and extend up to high eccentricity values, remaining linearly stable. The linear dependence of the eccentricities along the families, which is mentioned by Giuppone et al. (2010), seems to hold only for eccentricities less that $0.6$. For larger eccentricities, the characteristic curves bend and tend to the diagonal $e_1=e_2$ as $e_i\rightarrow 1$. Along the asymmetric families the phases $\Delta\omega$ and $\Delta M$ vary. Their variation is presented in Fig. \ref{FigFamsA}b, where the families are parametrized by the variable $e_1$ that defines the horizontal axis. The families start from $(\Delta\omega, \Delta M)$=$(240^\circ,180^\circ)$ (or, $(60^\circ,180^\circ)$ for their mirror image) and, as $e_i\rightarrow 1$, end to  $(180^\circ,0^\circ)$, i.e. to a symmetric configuration. The existence of such asymmetric orbits in the general three body problem can be explained by the mass continuation of asymmetric orbits of the restricted problem similarly to the 2/1 resonance studied in Voyatzis et al. (2009). We note that the restricted problem has various families of asymmetric orbits associated with the Lagrangian equilateral solutions (Zagouras et al. 1996; Papadakis and Rodi 2010).

\subsection{Stability regions around periodic orbits} \label{Sectionmaps}

\begin{figure}
\begin{center}
$\begin{array}{ccc}
\includegraphics[height=8cm]{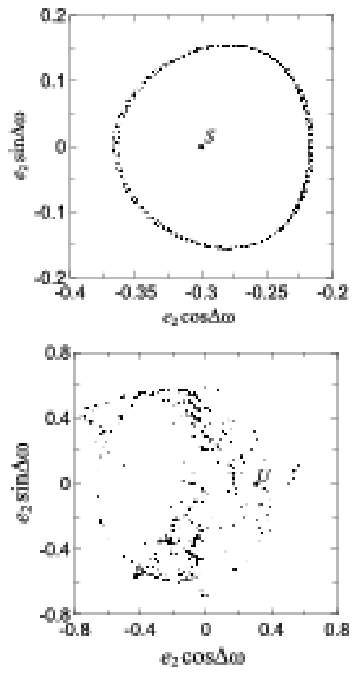} & \qquad &
\includegraphics[height=8cm]{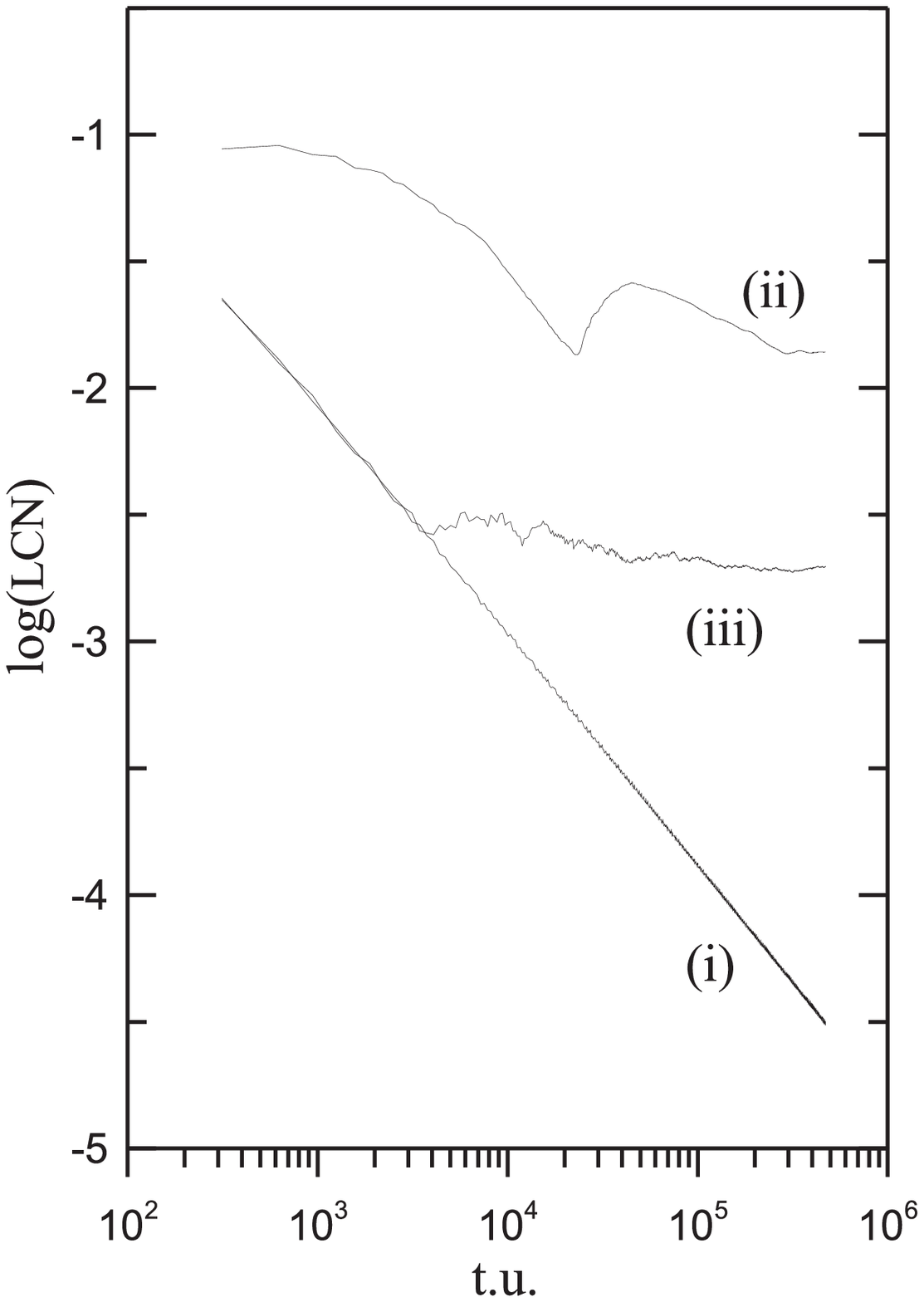}\\
\textnormal{(a)} & \qquad & \textnormal{(b)}
\end{array} $
\end{center}
\caption{{\bf a} Poincar\'e maps near a periodic orbit of the stable family $(S)$ (upper panel) and of an orbit which starts very close to the unstable periodic orbit $(U$) (bottom panel), for $m_1=0.001$, $m_2=0.001$. {\bf b} The evolution of the LCN along the stable and the unstable orbit of panel a (curves (i) and (ii), respectively). The evolution of the LCN along a relatively weakly chaotic orbit is also shown (curve (iii)). The initial conditions  are $e_1=e_2=0.3$, $\omega_2=0^\circ$ $M_1=180^\circ$, $M_2=0^\circ$ in all cases and  $\omega_1=150^\circ,\;0^\circ$ and $138^\circ$ for the  cases (i)-(iii), respectively.}
\label{FigPncLcn}
\end{figure}

It is well known that stable periodic orbits are surrounded by invariant tori, which correspond to longterm regular evolution. On the other hand, starting nearby an unstable periodic orbit we obtain in most cases chaotic evolution. Particularly, in the framework of the general three body problem, chaotic motion generally leads the planetary system to disruption. An example is shown in the two panels of Fig. \ref{FigPncLcn}a, where we present two typical orbits by using the Poincar\'e map on the plane $(e_2 \cos\Delta\omega,e_2 \sin\Delta\omega)$. The first one starts with initial conditions close to a stable periodic orbit in Fig. \ref{FigFamsS}, for $\rho=1$ ($m_1=0.001$, $m_2=0.001$), whose fixed point on the map is indicated by a filled circle. Its evolution shows a regular distribution of points that encompass the stable fixed point. The second orbit starts with initial conditions that correspond to an unstable periodic orbit for $\rho=1$. After a few revolutions the orbit deviates from the unstable periodic orbit and shows an irregular distribution of points on the Poincar\'e section. Finally, after a few thousand revolutions, the planets suffer a close encounter and the system leaves the 1/1 resonance. 

The regular or the chaotic character of the planetary evolution can be studied by various chaotic indicators (see e.g. Voyatzis, 2008). In Fig. \ref{FigPncLcn}b we show the evolution of the Lyapunov characteristic number (LCN) of the stable and the unstable orbits of Fig. \ref{FigPncLcn}a (curves (i) and (ii), respectively). Chaos and order is easily distinguished in this case. The curve (iii) corresponds to a relatively weak chaotic evolution where the planets show a slow irregular diffusion and the system will be disrupted in some hundred thousands revolutions. In the particular systems we study, since the 1/1 resonance includes close encounters, most of the orbits starting in the resonance are strongly chaotic, while orbits like case (iii) are rare.  

In order to study the range of stability around periodic orbits we compute dynamical stability maps constructed by using 2D grids of initial conditions around periodic orbits and computing the LCN at time intervals that correspond approximately to $10^5$ planetary revolutions for a regular orbit. If during this integration time interval a close encounter occurs we set LCN=1. For the case of orbits of planetary type we consider initial conditions given in orbital elements. However, for orbits of satellite type, where the osculating orbital elements with respect to the star become meaningless, the initial conditions are given in variables of the rotating frame.

\begin{figure}
\begin{center}
\includegraphics[width=11cm]{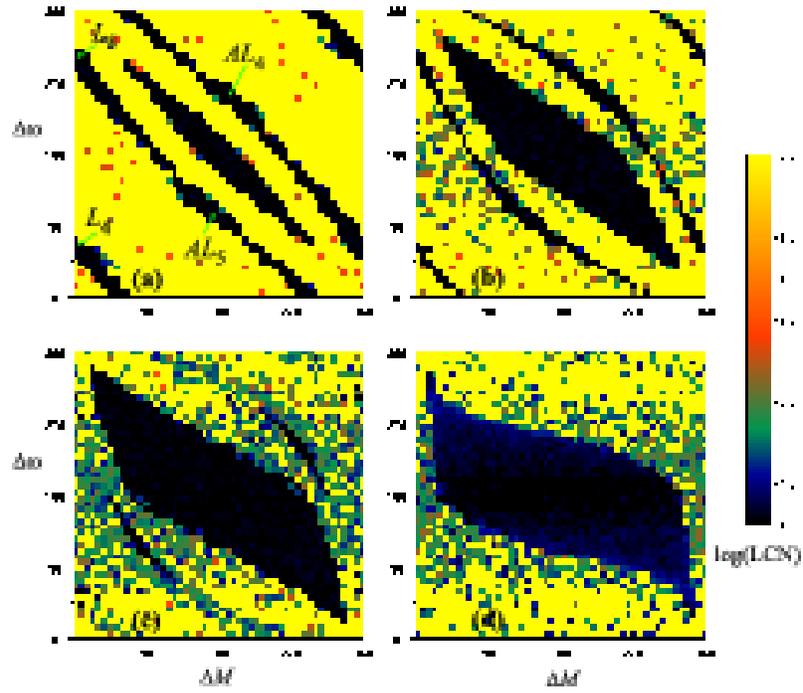} 
\end{center}
\caption{LCN dynamical maps ($50\times 50$) around the symmetric stable periodic orbit $(\Delta M,\Delta \omega)$=$(180^\circ,180^\circ)$ at {\bf a} $e_i=0.2$, {\bf b} $e_i=0.4$, {\bf c} $e_i=0.57$, {\bf d} $e_i=0.8$ ($i=1,2$).}
\label{FigPhaseMaps}
\end{figure}

We consider the planetary part of the stable family in Fig. \ref{FigFamsS} for $\rho=1$ ($m_1=m_2=0.001$), where for all of its periodic orbits it is $a_1\approx a_2$, $e_1\approx e_2$, $\Delta\omega=180^\circ$ and $\Delta M=180^\circ$. We select four typical periodic orbits along the family to study the stability region around them, at the points $e_i=0.2$, $e_i=0.4$, $e_i=0.57$ and $e_i=0.8$ ($i=1,2$). The corresponding four dynamical maps are presented in Fig. \ref{FigPhaseMaps}. For the computation of the dynamical maps we fixed the initial conditions $a_i$ and $e_i$ at the periodic orbit and varied the angles $\Delta\omega$ and $\Delta M$. $50\times 50$ grids are formed by considering values $\Delta\omega$ and $\Delta M$ in the interval $[0,360]$. The symmetric periodic orbits are located in the center of the maps, while the asymmetric periodic orbits of the families $AL_4$ and $AL_5$ correspond at the indicated orbits. Around the symmetric periodic orbit at $e_1$=$e_2$=$0.2$ (panel a), which is located on the planetary part of the family close to the cusp (see family for $\rho=1$ in Fig. 3a), we find a thin island of stability (dark colored region) with quasiperiodic trajectories. This island is also surrounded by a stability strip, which is represented by the four dark zones (note the toroidal mod $2\pi$ periodicity of the maps). This region includes the asymmetric periodic orbits and the Lagrangian equilateral solutions $L_4$ and $L_5$ (see also Giuppone et al. 2010). As we pass to larger eccentricities (see panel (b)) the size of the stability island around the symmetric periodic orbit increases while the width of its surrounding  thin stability region decreases. For $e_1$=$e_2$=$0.57$ only two small stability islands remain around $AL_4$ and $AL_5$, which disappear for larger eccentricities. The large stability region around the symmetric periodic orbit remains even for high eccentricities $e_1=e_2=0.8$.

\begin{figure}
\begin{center}
$\begin{array}{ccc}
\includegraphics[width=6cm]{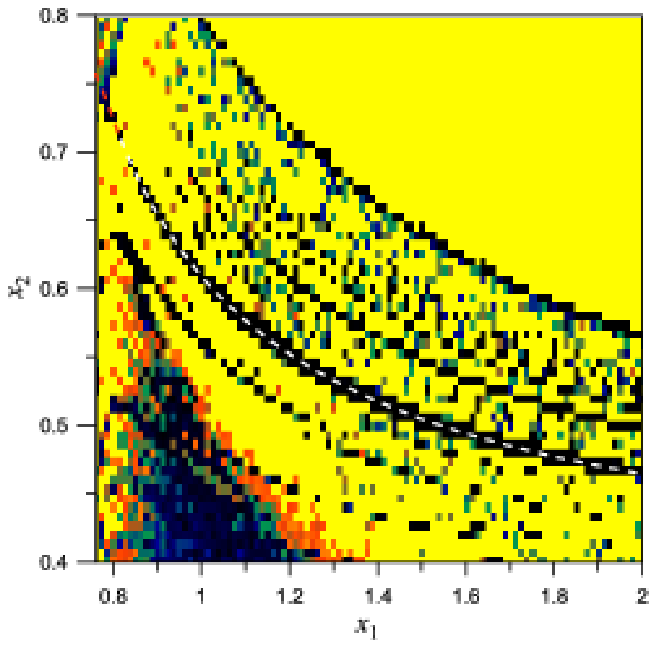} & \qquad &
\includegraphics[width=6cm]{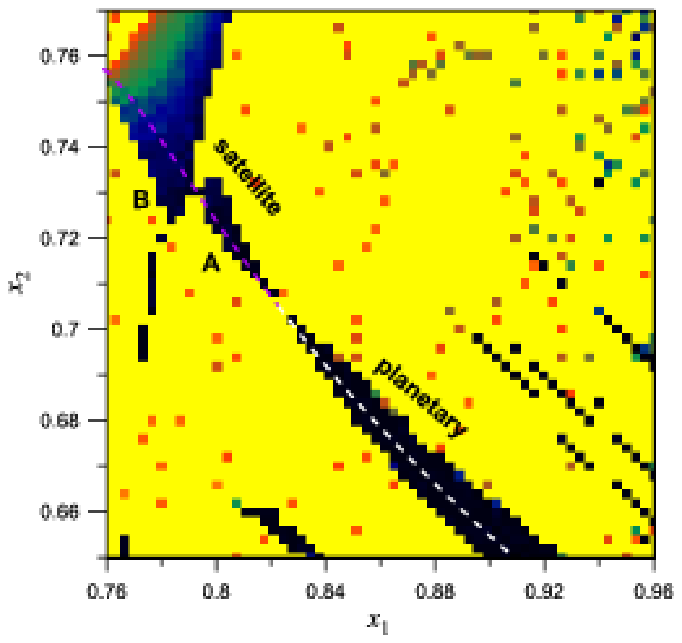}\\
\textnormal{(a)} & \qquad & \textnormal{(b)}
\end{array} $
\end{center}
\caption{ {\bf a} LCN dynamical map on the $x_1-x_2$ plane. The stable symmetric family in this plane is indicated by the dashed curve. {\bf b} A magnification of the map of panel (a) at the region where passage from planetary to satellite orbits occurs.}
\label{FigMapx1x2}
\end{figure}

In order to study how the stability regions are formed, as we pass from planetary to satellite orbits, we consider grids with initial conditions along the above stable family ($m_1=m_2=0.001$) which can be parametrized by using the variable $x_1$ of the rotating frame at the horizontal axis of the grid. Thus for each value of $x_1$ the orbits in the grid are fixed to the initial conditions of the corresponding periodic orbit and we vary another variable, as shown in the vertical axis of the grid. In Fig. \ref{FigMapx1x2}a we present such a grid with vertical axis the variable $x_2$ of the rotating frame. The projection of the stable family in this plane is also presented. We obtain that around the planetary part of the family a strip of regular orbits is formed. However, at the passage point from planetary to satellite orbits at $x_1\approx 0.82$ the stability strip breaks (see Fig. \ref{FigFamsS}). This is clearly shown in Fig. \ref{FigMapx1x2}b where the magnification of this region is presented. We observe, also, a second break of the stability zone at $x_1\approx 0.79$. So in the satellite part, we get two distinct regions (A and B) which show compact domains of regular orbits.

\begin{figure}
\begin{center}
$\begin{array}{ccc}
\includegraphics[width=6cm]{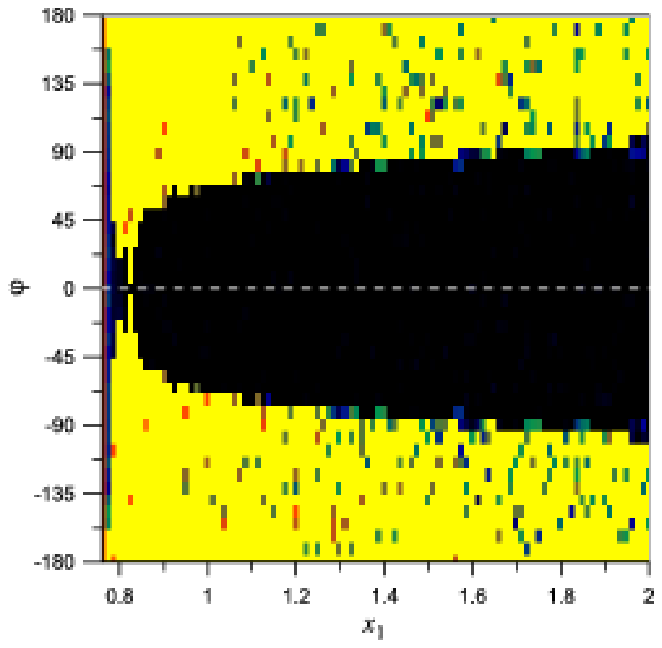} & \qquad &
\includegraphics[width=6cm]{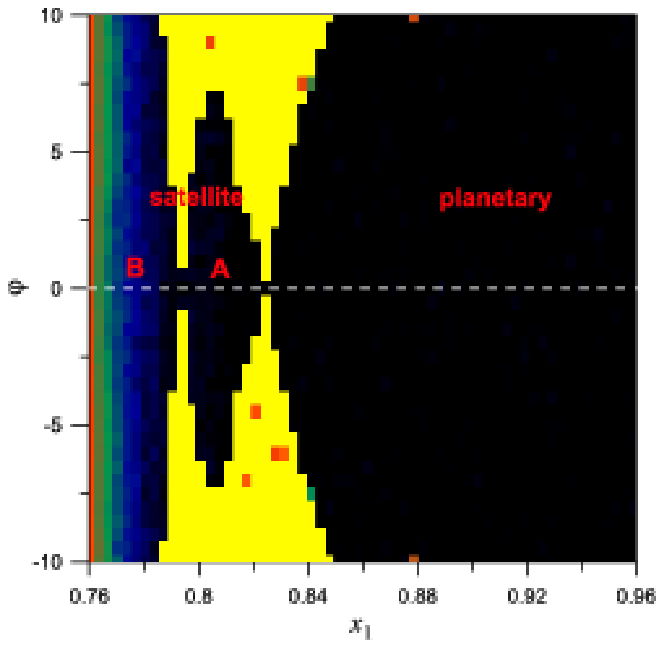}\\
\textnormal{(a)} & \qquad & \textnormal{(b)}
\end{array} $
\end{center}
\caption{ {\bf a} LCN dynamical map on the $x1-\phi$ plane. The stable family of symmetric periodic orbits in this plane is indicated by the dashed curve. {\bf b} A magnification of the map of panel (a) at the region where passage from planetary to satellite orbits occur.}
\label{FigMapxav}
\end{figure}

Another dynamical map that illustrates the stability regions close to the transition point from planetary to satellite orbits is given in Fig. \ref{FigMapxav}. Now the vertical axis presents the angle $\phi$ between the velocity vector $\mathbf{v}=(\dot x_2,\dot y_2)$ and the vertical axis $Oy$ of the rotating frame. The norm of the vector is fixed to the value that corresponds to the periodic orbit. Note that for $\phi=0$ we get the periodic orbits of the stable family. The dynamical map shows that along the planetary part of the family, we get a relatively large region of stability which is extended approximately in the domain $-60^\circ<\phi<-60^\circ$. As we approach the cusp of the family, the width of the stability region decreases rapidly and is cut at the cusp. After this point ($x_1<0.82$) we get the stability region of the satellite orbits which shows the two distinct regions A and B consisting of regular  orbits.
We should remind at this point that the family ends at a collision point of the planets at $x_1\approx 0.76$, which indicates the end of the domain B. Thus the strong chaos close to this value is due to the close encounters of the planets and the possible errors of the numerical integration of the close to collision orbits. 
The above results were obtained for $m_1=m_2=0.001$ ($\rho=1$), but we checked that the same properties exist for any other mass ratio.

%%%%%%%%%%%%%%%%%%%%%%%%%%%%%%%%%%%%%%%%%%%%%%%%%%%%%%%%%%%%%
%%% SECTION 4
%%%%%%%%%%%%%%%%%%%%%%%%%%%%%%%%%%%%%%%%%%%%%%%%%%%%%%%%%%%%%

\section{Evolution of the system under non conservative forces} \label{NonConEvol}

We start with a planetary system with two planets, with large eccentricities, which is trapped in a 1/1 stable resonant periodic orbit. Its position in the phase space is a point on the planetary part of the stable family. If no external forces act to the system, this planetary system will stay in its initial condition for all $t$, i.e. it will be represented by a {\it fixed point} in the phase space. We assume now that, in addition to the gravitational forces, the dissipative force given by Eq. \ref{dis-law} acts also on the two planets. This latter system is no longer periodic, but is expected to evolve in time.  Our purpose is to study the long term behavior of such a system by following its evolution in phase space. The results will be presented in projections in coordinate planes and mainly in the eccentricity plane $e_1 - e_2$.

We checked many cases, with different total planetary masses $m_1+m_2$ and different planetary ratios $m_1/m_2$. Note that due to the fact that we are in the 1/1 resonance, which implies almost equal semimajor axes, $a_1\approx a_2$, we can study only the case $m_1/m_2\leq 1$. In all cases the long term evolution is the same, so we present in the following two typical cases, $m_1=0.001$, $m_2=0.001$ and mainly the case $m_1=0.0001$, $m_2=0.0010$.

\begin{figure}
\begin{center}
$\begin{array}{cccc}
\includegraphics[width=5cm,height=5cm]{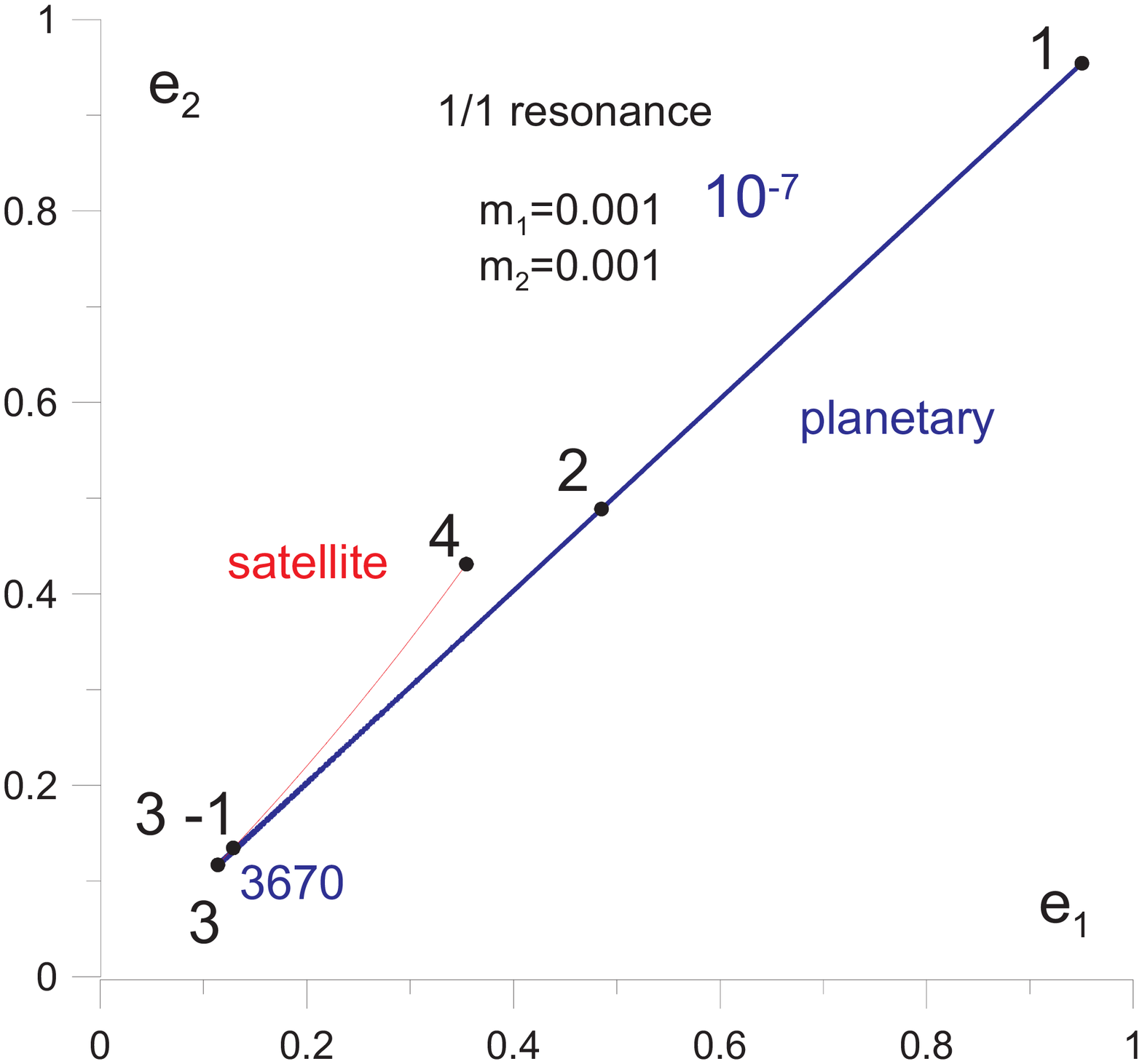} & \qquad&
\includegraphics[width=5cm,height=5cm]{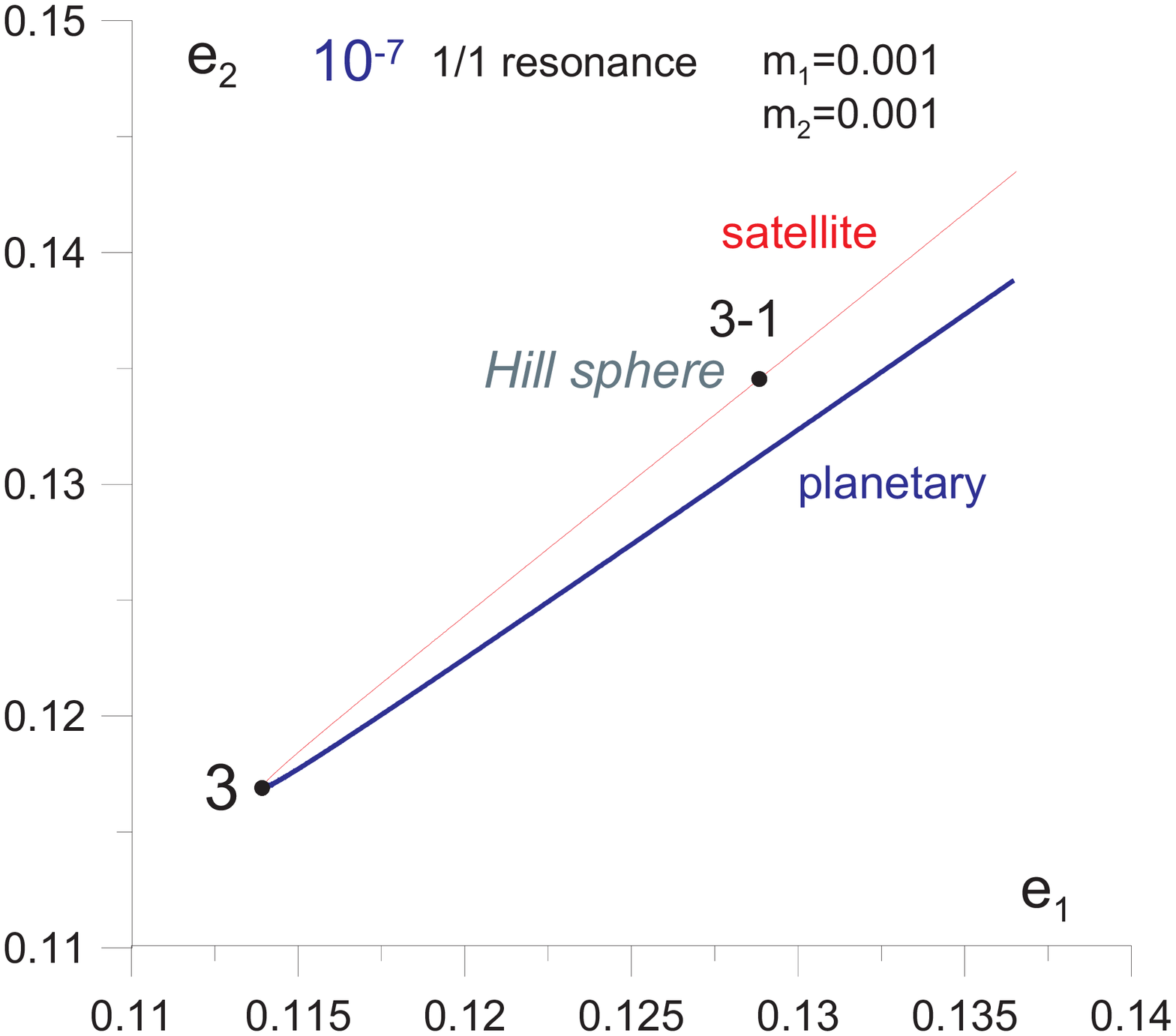}& \\
\textnormal{(a)}  & \qquad& \textnormal{(b)}  
\end{array}$ 
\end{center}
\caption{{\bf a} The evolution, in the eccentricity space, of the system under the dissipative force (Eq. \ref{dis-law}), for $n=7$, $m_1=0.00100$, $m_2=0.00100$ and for the initial conditions \ref{init_1-1}. The starting point is {\it on} the family. The whole curve represents the evolution of {\it one} orbit. {\bf b} Detail close to the transition from planetary to satellite orbits. The family of panel (a) is the same as the family of Fig. \ref{FigFamsS}a, for $\rho=1$.} 
\label{e12-dis}
\end{figure}
\subsection{Typical cases of evolution}

In Fig. \ref{e12-dis} we present the evolution of a planetary system that starts on the planetary part of the stable family of 1/1 resonant periodic orbits, with initial conditions
\begin{equation}
\begin{array}{ll}
a_1 = 8.59425 \ \ & a_2 = 8.56898\\
e_1 = 0.95077 \ \ &  e_2 = 0.95486\\
\omega_1 = 0 \ \ &\omega_2=\pi\\
M_1 = \pi & M_2 = 0
\end{array}
\label{init_1-1}
\end{equation}
and $m_1=0.00100$, $m_2=0.00100$, that correspond to an exact periodic orbit. We assume that the dissipative force that acts to the planets is given by Eq. \ref{dis-law} with exponent $n=7$. To have a better physical understanding, we present in Fig.\ref{e12-dis}a  the evolution of the system in the eccentricity space. The system starts with high eccentricities and follows the planetary part of the family of resonant periodic orbits, with decreasing values of the eccentricities. This evolution continues along the family up to the {\it orbit 3}, which is the transition point from planetary to satellite orbits.  We remark that the apparent ``cusp'' does not imply any discontinuity, as explained in section \ref{SectionFams}. In Fig. \ref{e12-dis}b we present the detail of panel (a) in the vicinity of the transition point. The {\it orbit 3} is at the edge of the ``cusp'' and the {\it orbit 3-1} corresponds to the Hill sphere, as explained bellow. From this point on, the system evolves along the {\it satellite part} of the family and ends up to a close binary of $P_1$ around $P_2$ (planet-satellite system) whose center of mass revolves around the star in an almost circular orbit. 

\begin{figure}
\begin{center}
$\begin{array}{ccc}
\includegraphics[width=5cm,height=5cm]{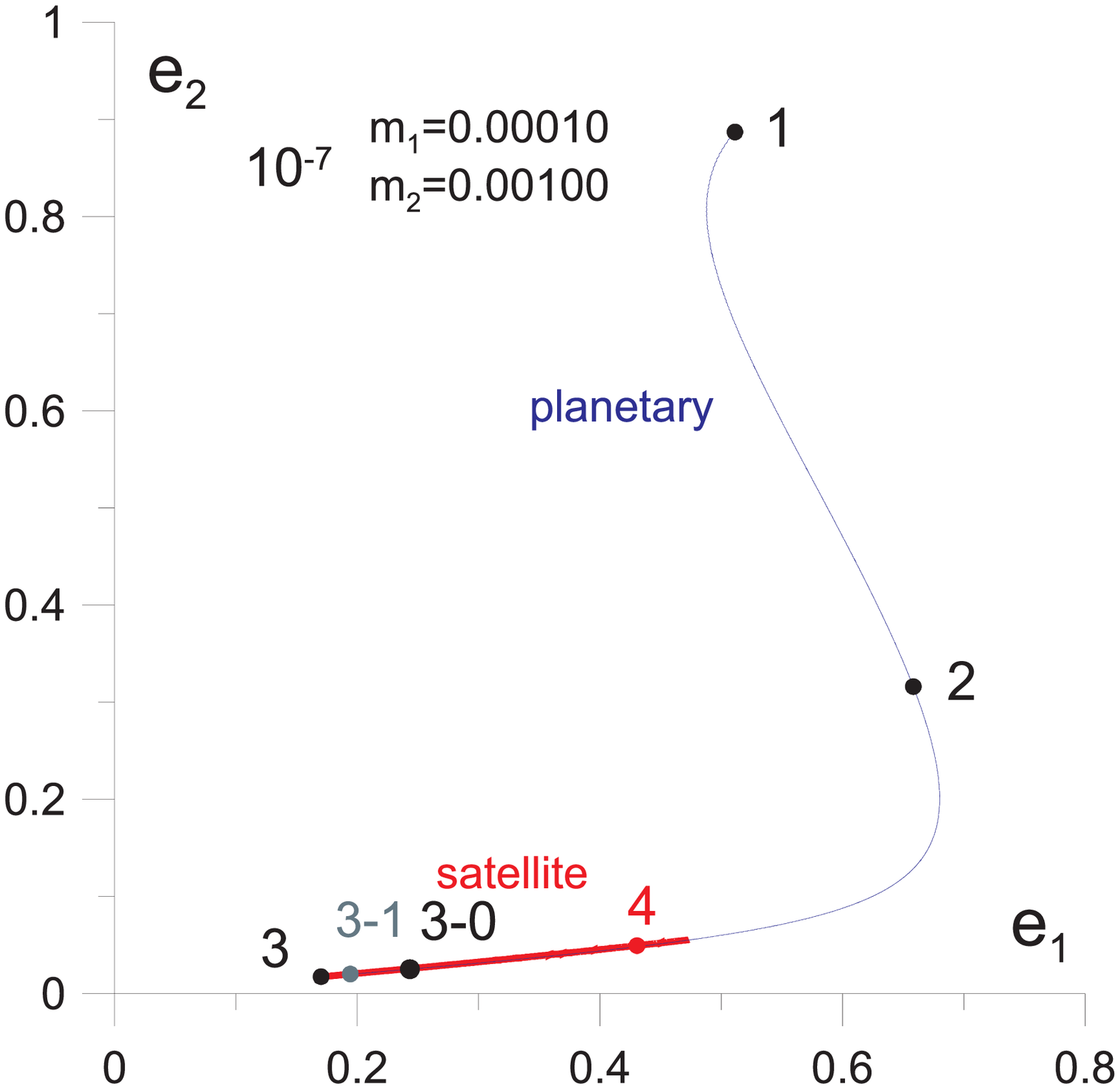} & \qquad&
\includegraphics[width=5cm,height=5cm]{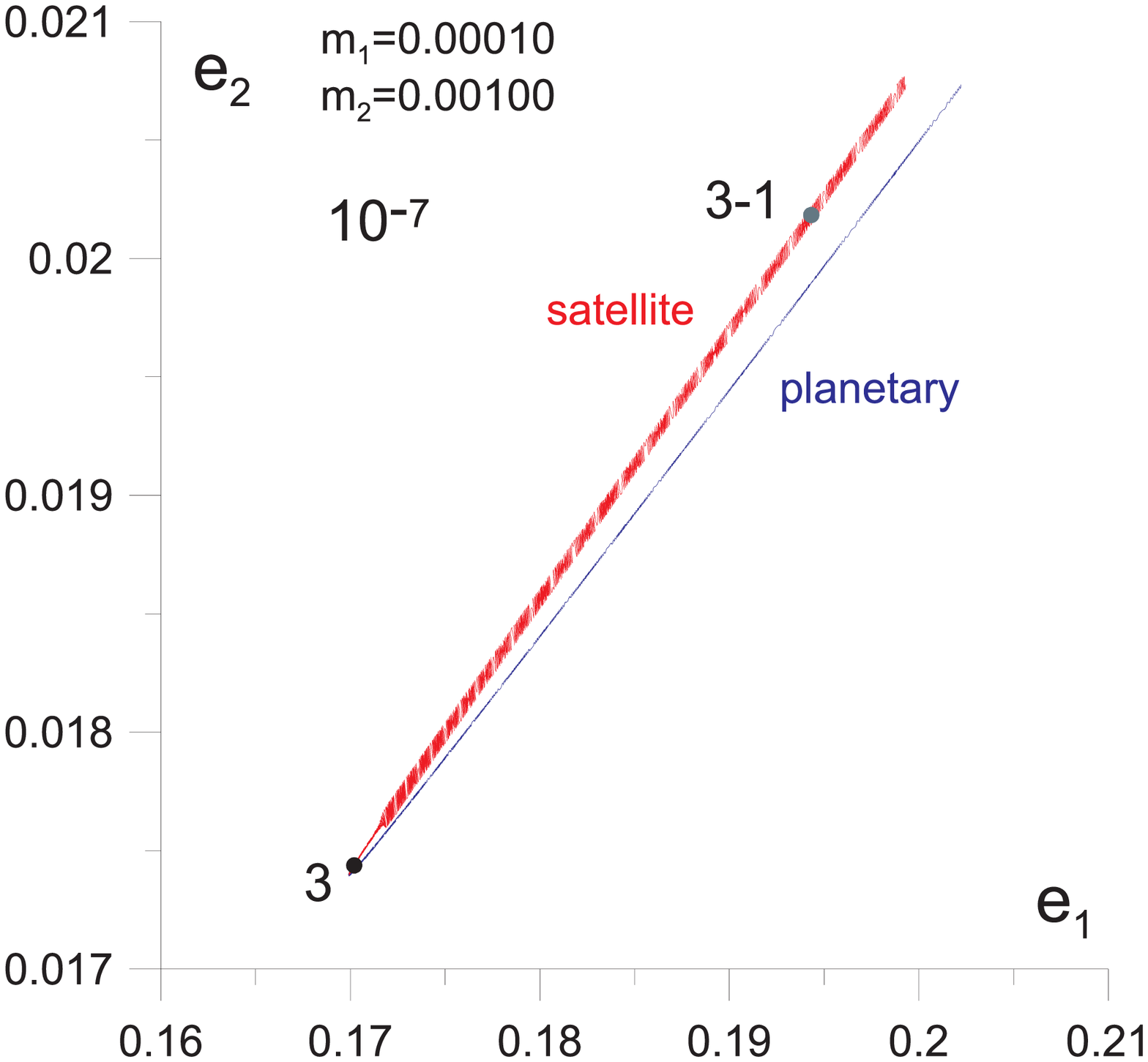} \\  
\textnormal{(a)} & \qquad& \textnormal{(b)} 
\end{array}$
\end{center}
\caption{The same as in Fig. \ref{e12-dis}a,b but for $m_1=0.00010$, $m_2=0.00100$ and for the initial conditions \ref{init_1-2}. The {\it orbit 3-0} is on the planetary part and the {\it orbit 3-1} is on the satellite part. The family of panel (a) is the same as the family of Fig. \ref{FigFamsS}a, for $\rho=10$.}  
\label{e12-dis1}
\end{figure}

In Fig. \ref{e12-dis} we studied the case where the two planetary masses are equal. We repeat now this work by studying  the case where the planetary masses are different. A typical case is shown in Fig. \ref{e12-dis1} for the masses $m_1=0.00010$, $m_2=0.00100$ and starting from a periodic orbit with initial conditions
\begin{equation}
\begin{array}{ll}
a_1 = 2.27306 \ \ & a_2 = 2.26067\\
e_1 = 0.51134 \ \ &  e_2 = 0.8869\\
\omega_1 = 0 \ \ &\omega_2=\pi\\
M_1 = \pi & M_2 = 0
\end{array}
\label{init_1-2}
\end{equation}
We checked that in all other cases, with different masses, the evolution is qualitatively the same. 

\begin{figure}
\begin{center}
$\begin{array}{ccccc}
\includegraphics[width=3cm,height=3cm]{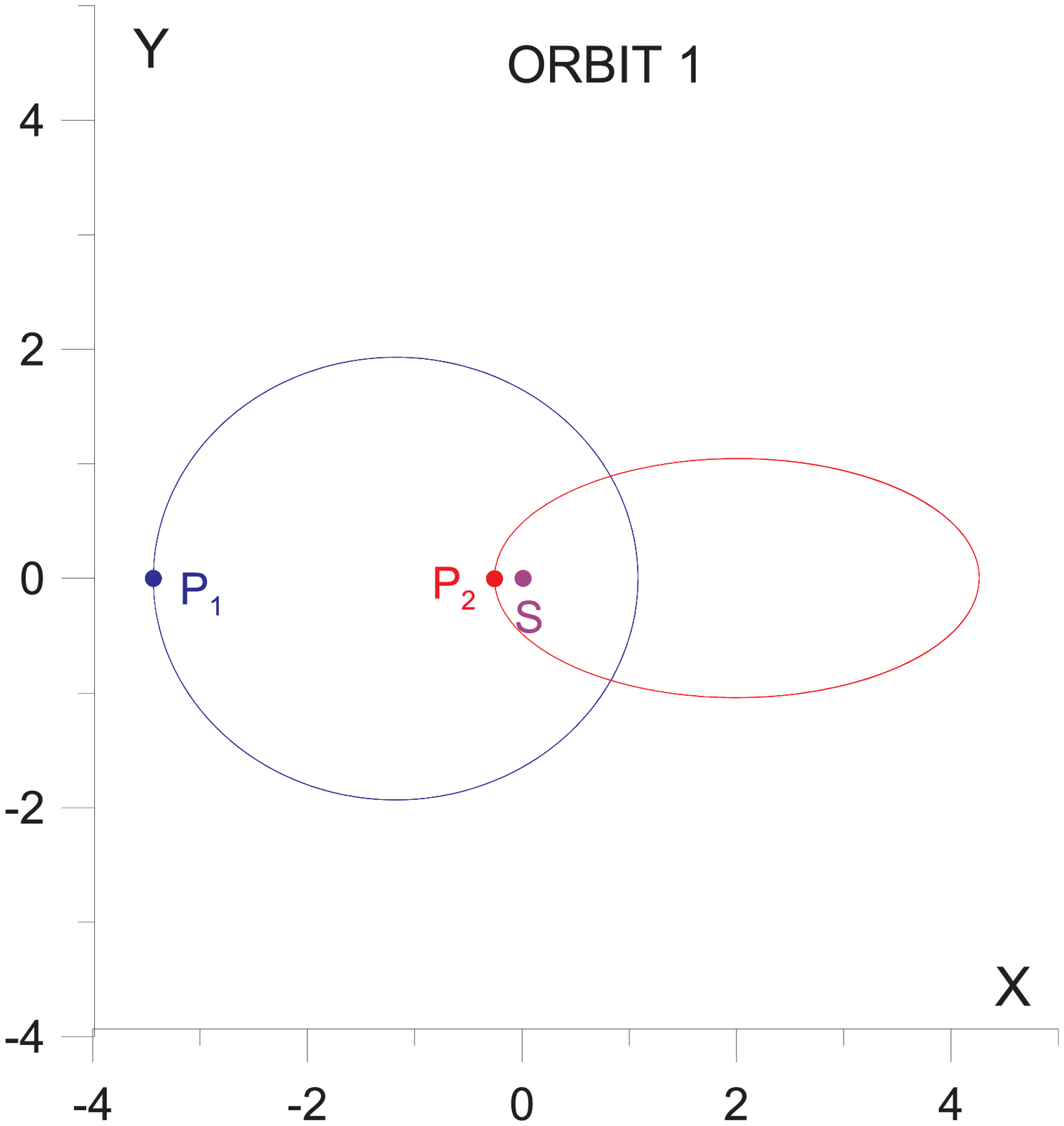} & 
\includegraphics[width=3cm,height=3cm]{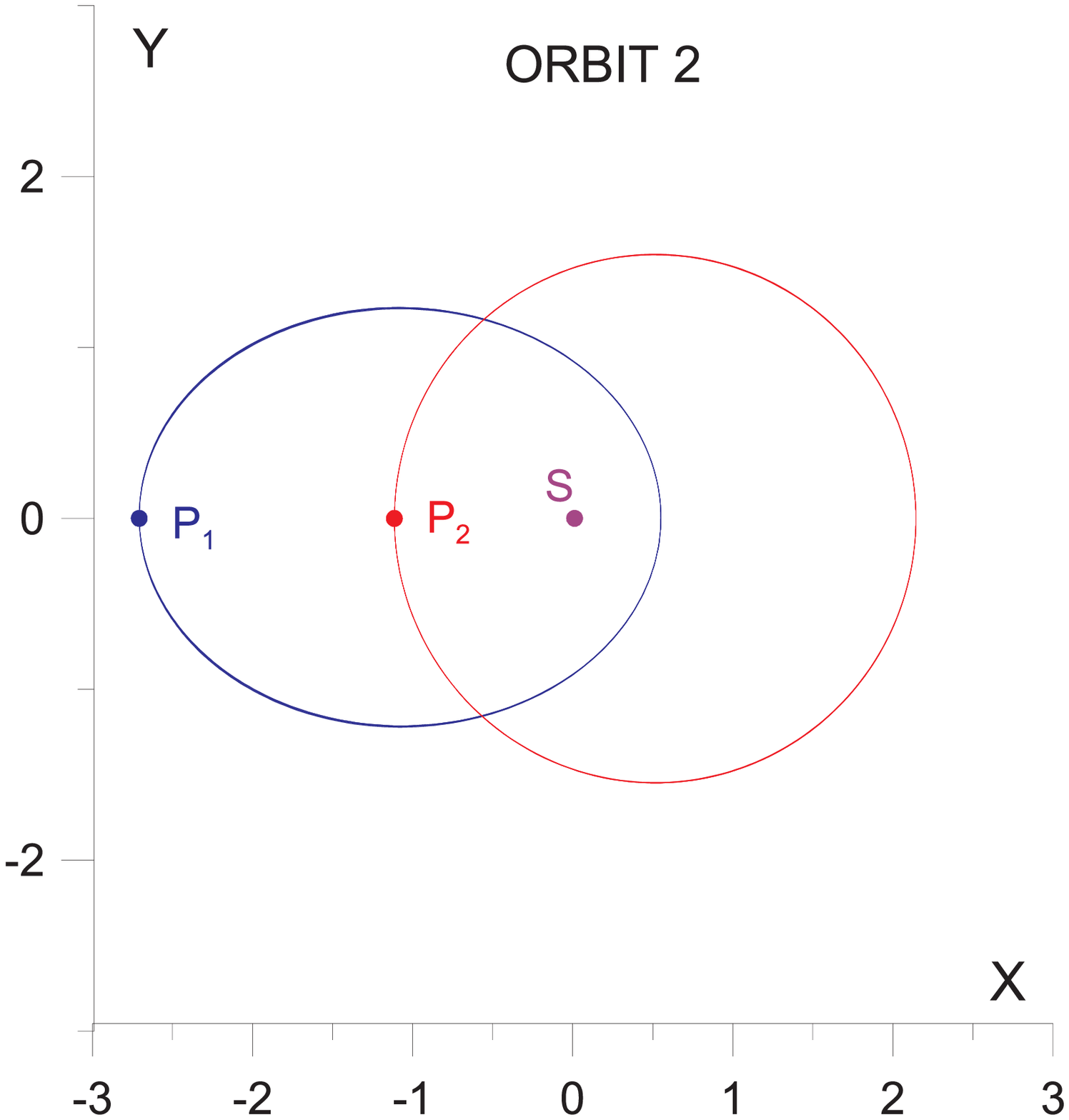} & 
\includegraphics[width=3cm,height=3cm]{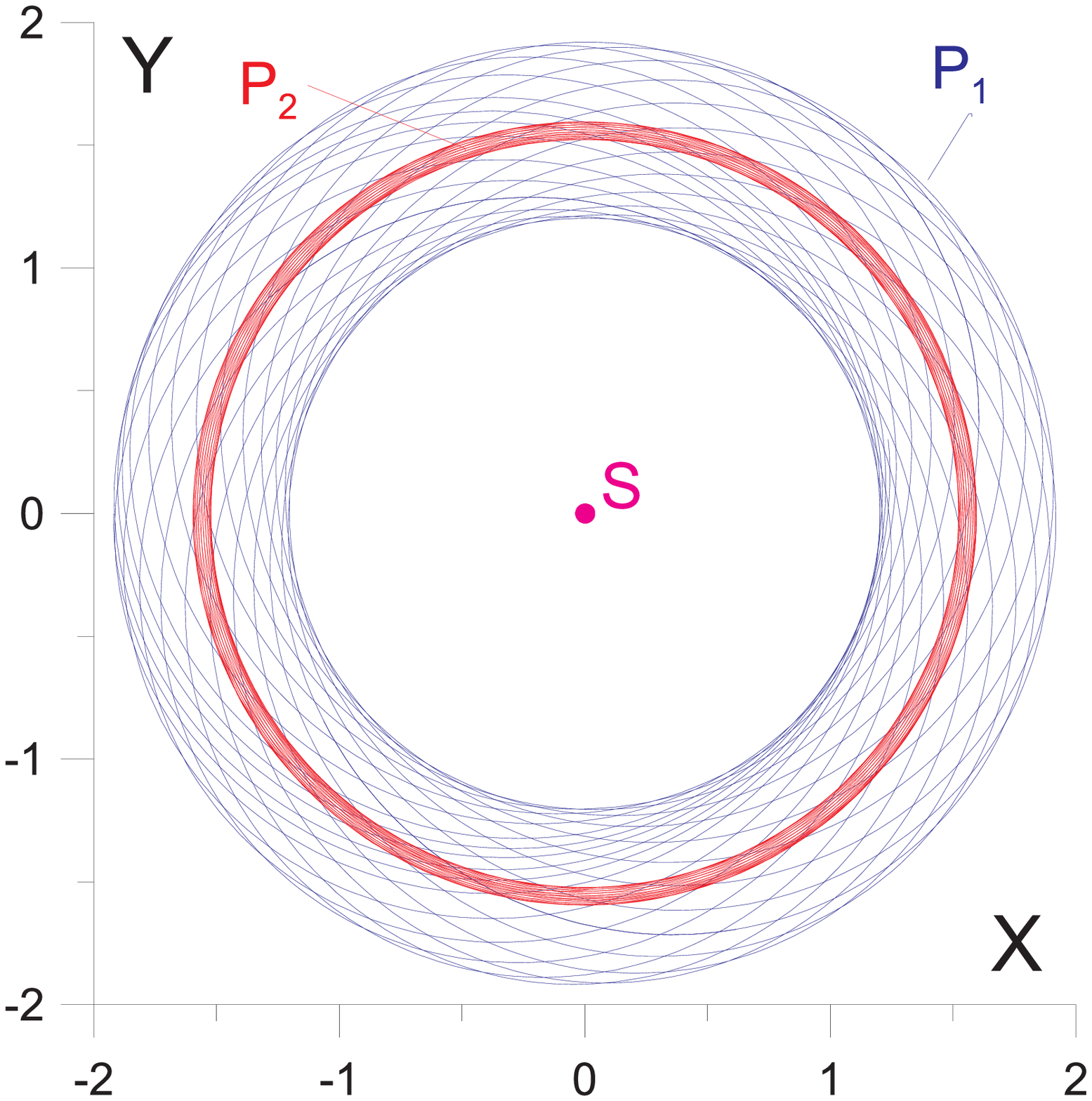} & 
\includegraphics[width=3cm,height=3cm]{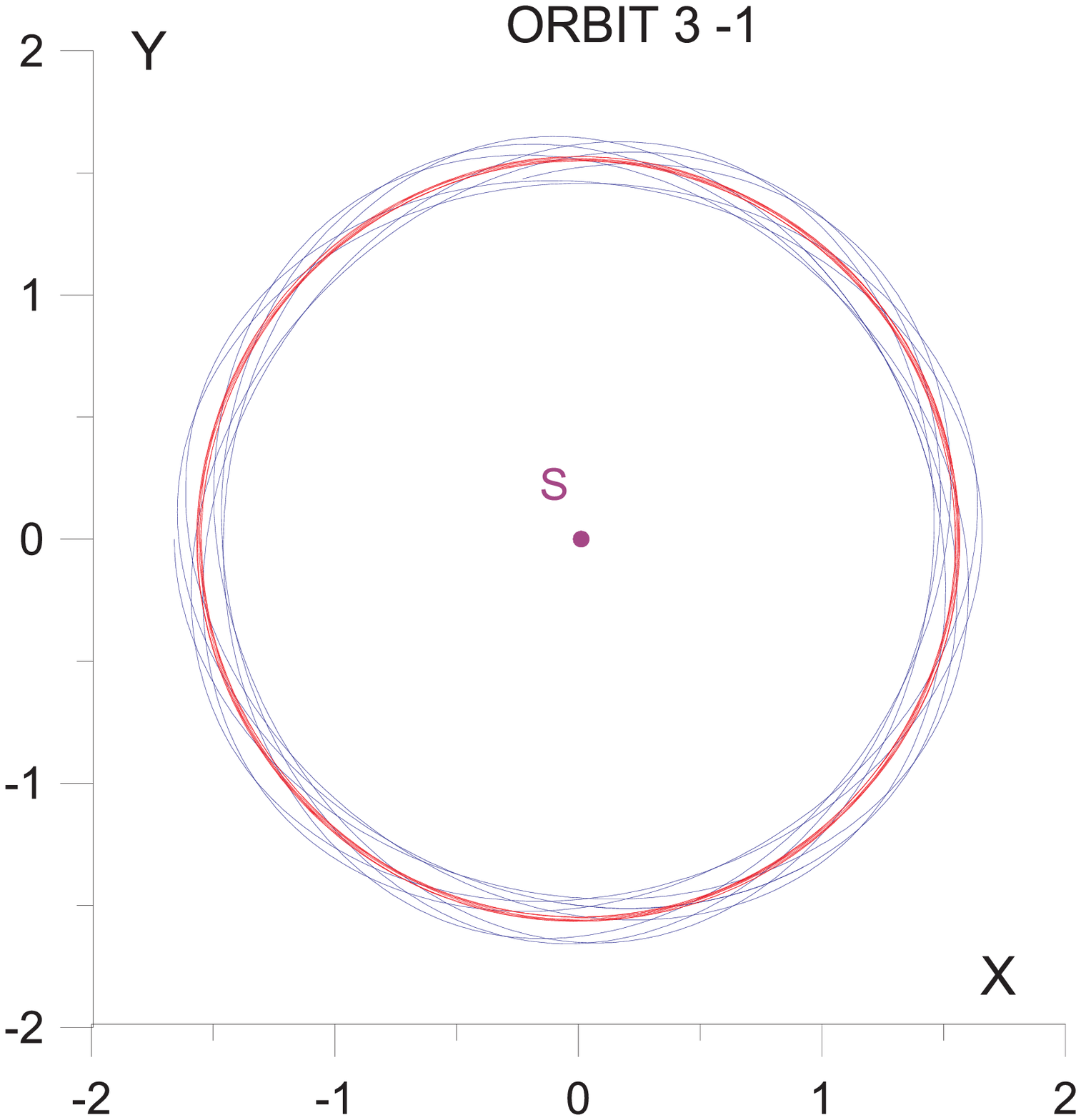} & \\
\textnormal{(a)} &\textnormal{(b)}&\textnormal{(c)} &\textnormal{(d)}
\end{array}$ 
\end{center}
\caption{Four typical orbits along the continuous evolution presented in Figs. \ref{e12-dis1}a,b, at the corresponding numbers: {\bf a} The {\it orbit 1}: large eccentricities. {\bf b} The {\it orbit 2}: intermediate eccentricities. {\bf c} The {\it orbit 3-0} on the planetary part, just before the transition point. The motion is close to a satellite orbit.  {\bf d} The {\it orbit 3-1} on the satellite part, just after the transition point.}  
\label{orbits123}
\end{figure}
\begin{figure}
\begin{center}
$\begin{array}{cccc}
\includegraphics[width=3.2cm,height=3.2cm]{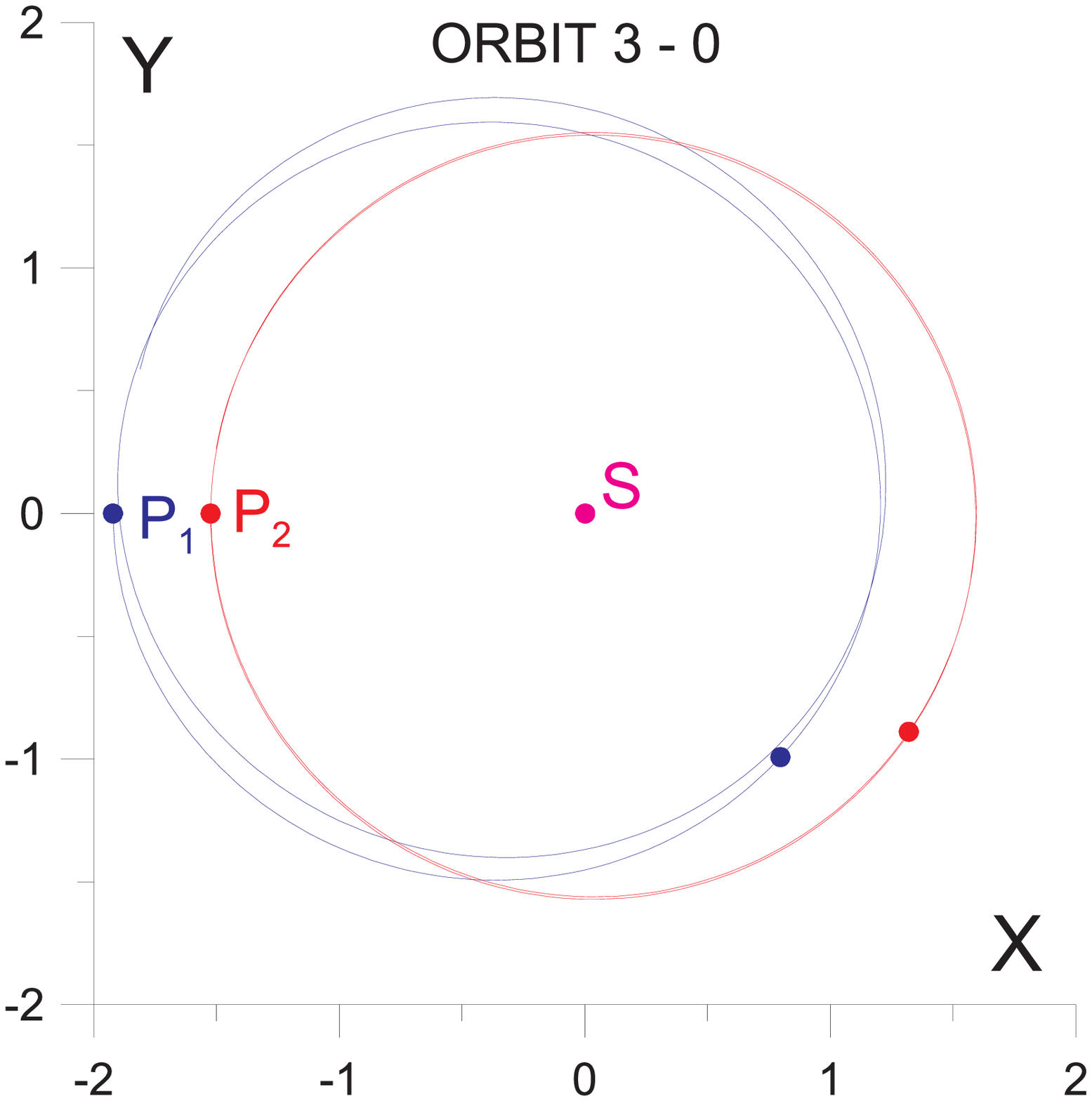} & 
\includegraphics[width=3.2cm,height=3.2cm]{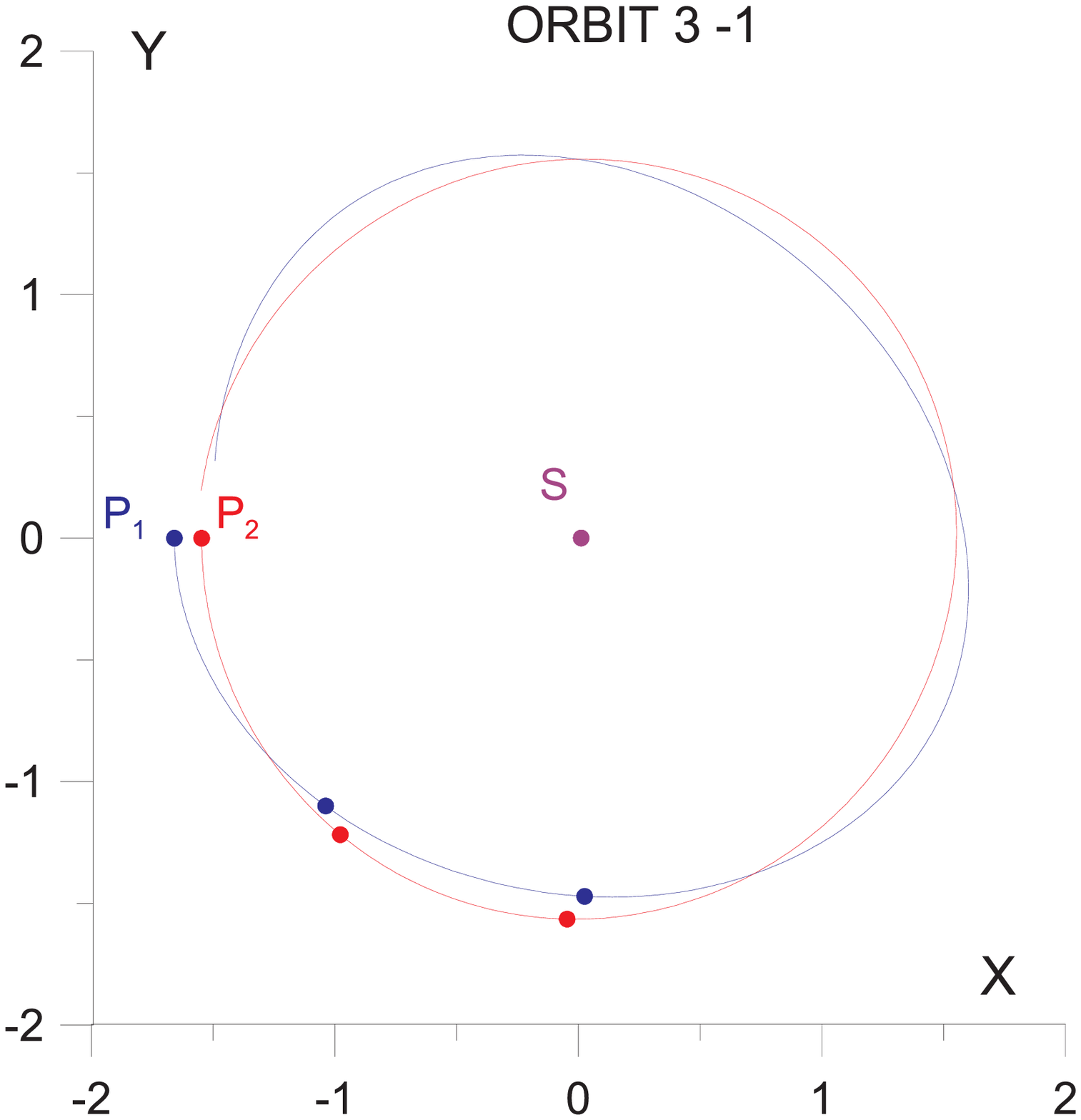} & 
\includegraphics[width=3.5cm,height=3.2cm]{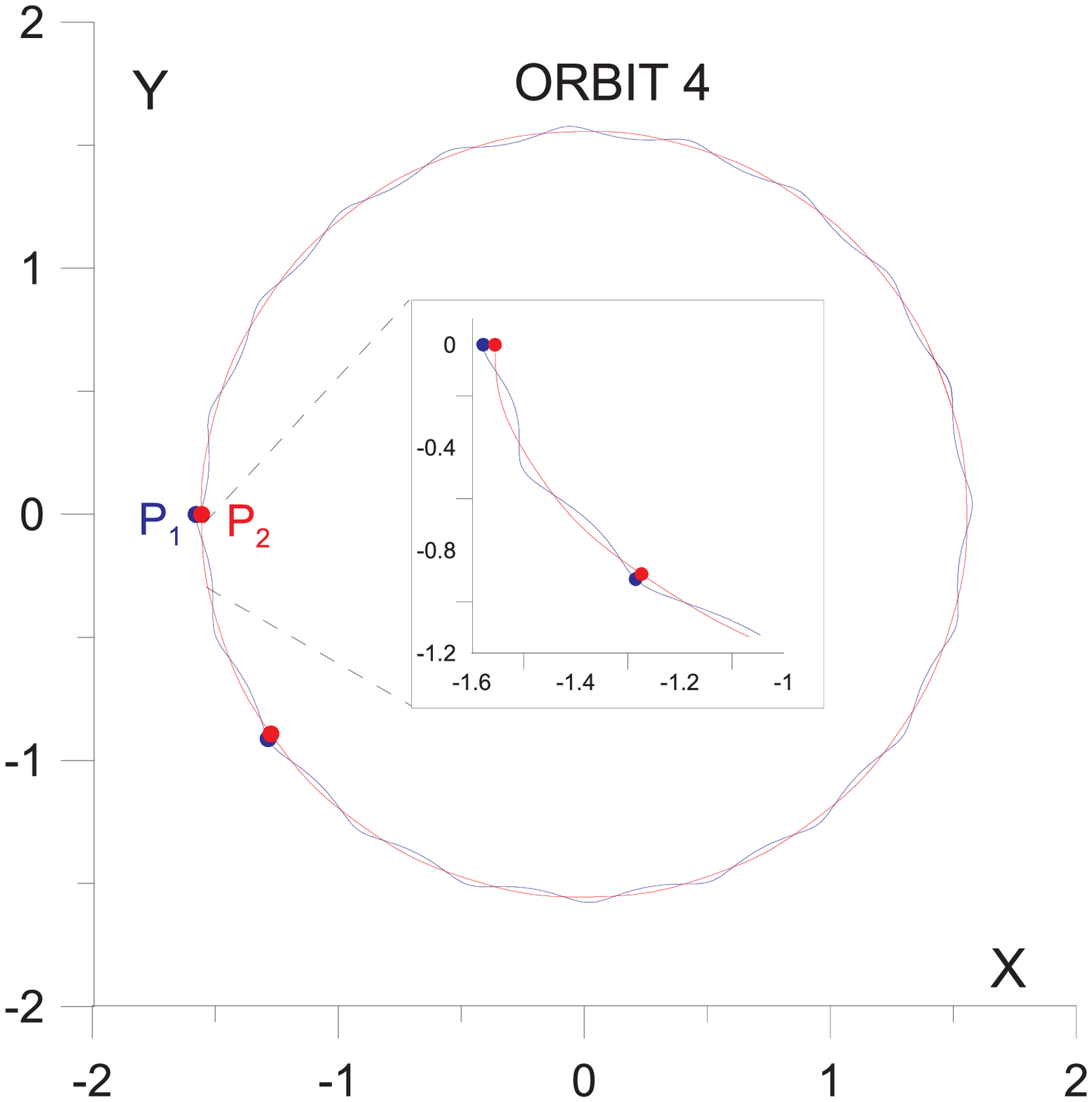} & \\
\textnormal{(a)} &\textnormal{(b)}&\textnormal{(c)}
\end{array}$ 
\end{center}
\caption{Detail of Fig. \ref{orbits123}: {\bf a} The {\it orbit 3-0}. The two planets move close to each other. {\bf b} The {\it orbit 3-1}. Trapping in a satellite orbit where $P_2$ moves in an almost circular orbit.  {\bf c} The {\it orbit 4}. Trapping in a close satellite orbit.}  
\label{orbits34}
\end{figure}
\begin{figure}
\begin{center}
$\begin{array}{cccc}
\includegraphics[width=3.5cm,height=3.2cm]{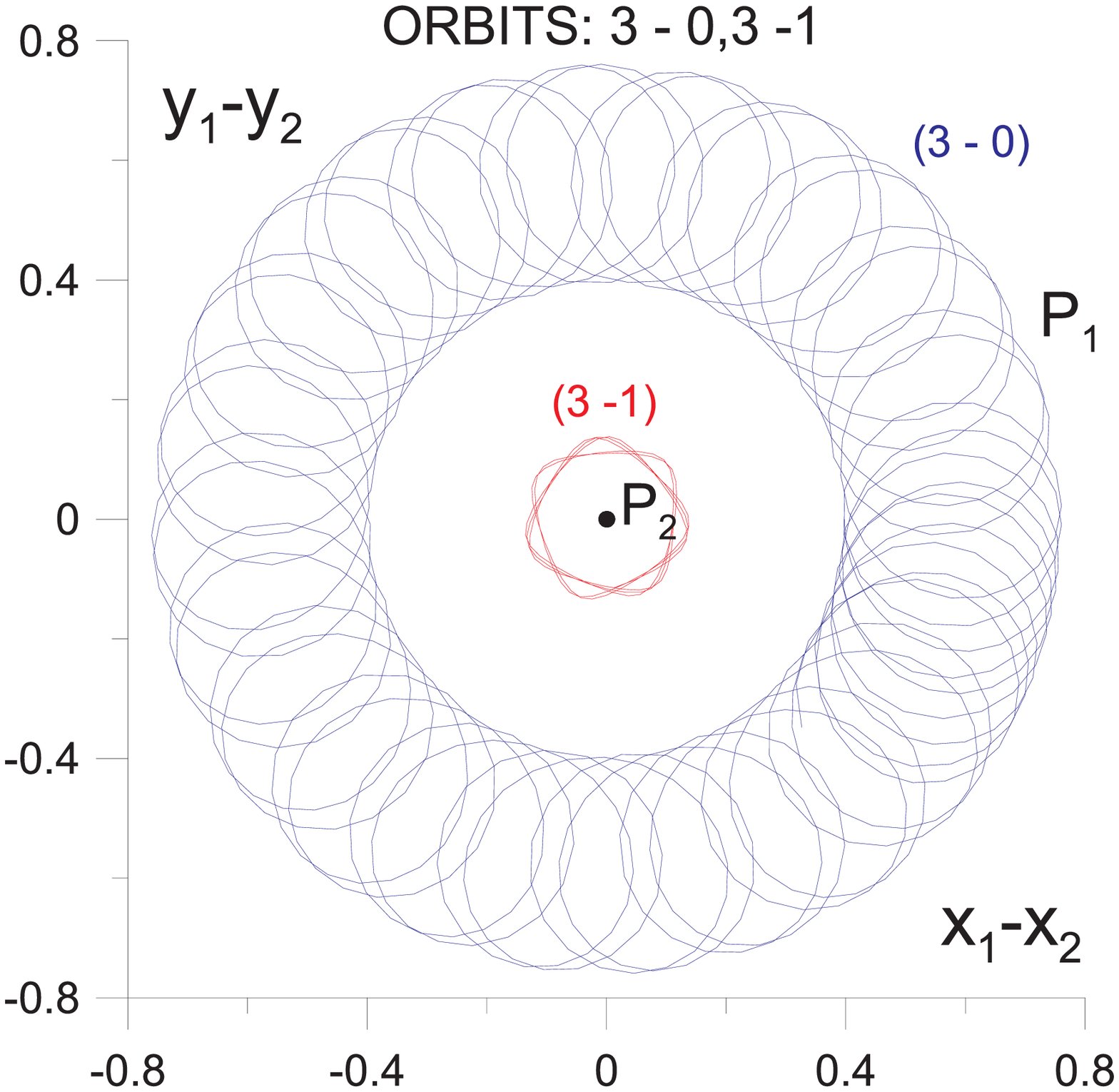} & 
\includegraphics[width=3.5cm,height=3.2cm]{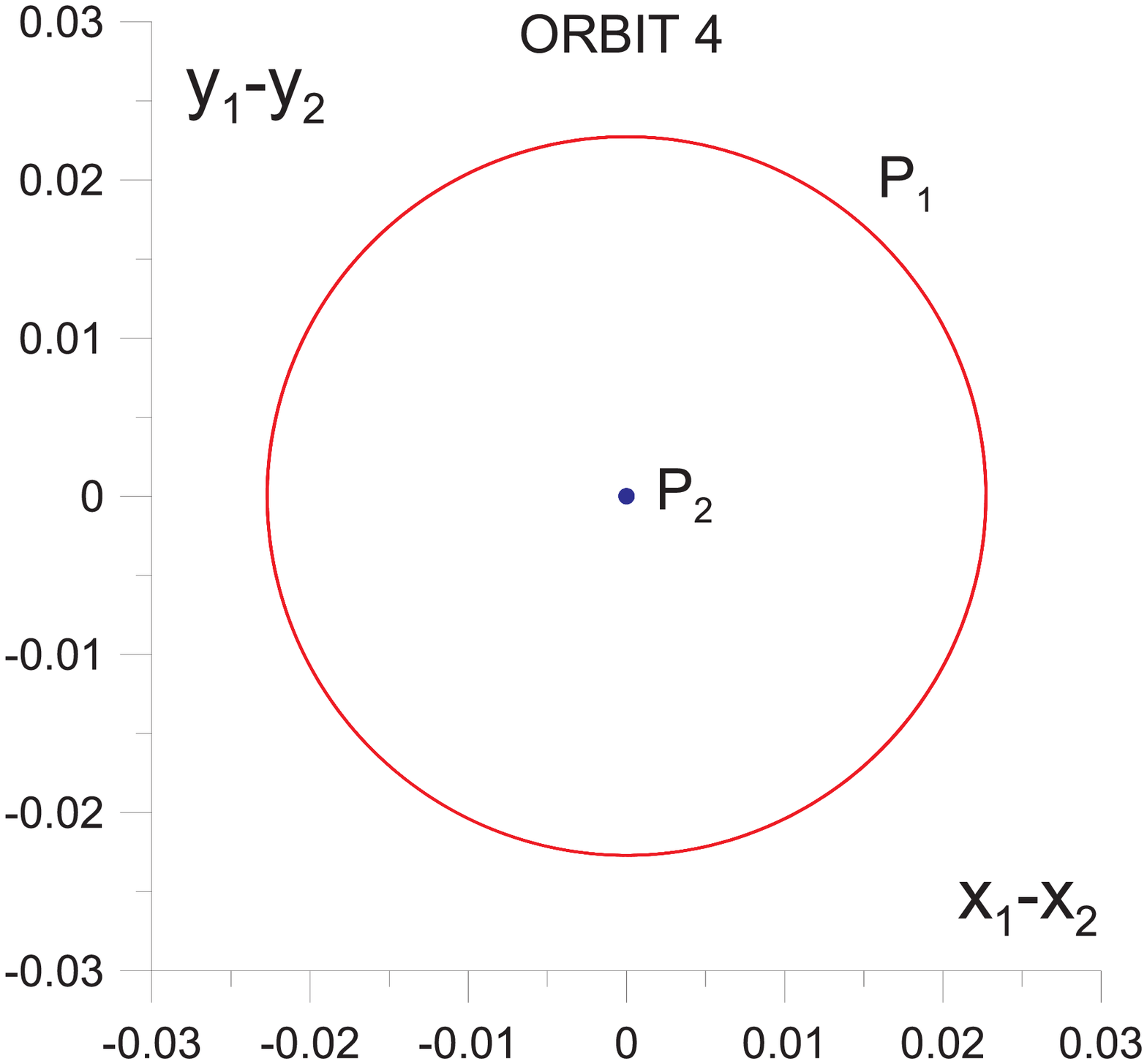}&
\includegraphics[width=3.5cm,height=3.2cm]{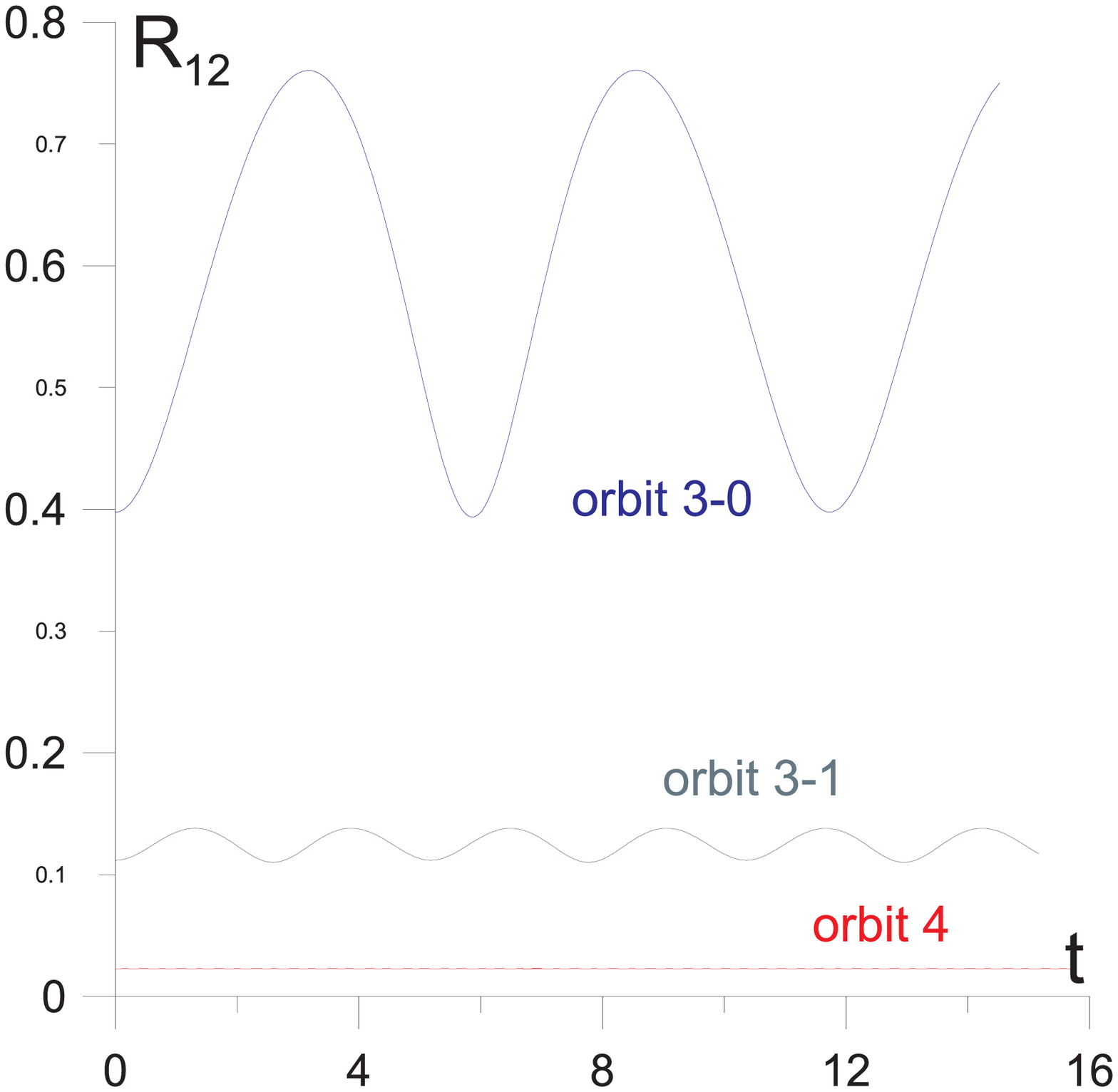}\\
\textnormal{(a)} & \textnormal{(b)} & \textnormal{(c)}
\end{array}$ 
\end{center}
\caption{The relative motion of $P_1$ around $P_2$ for {\bf a} The transition {\it orbits 3-0}, {\it 3-1} and {\bf b} The satellite {\it orbit 4} of Fig. \ref{orbits34}. This latter orbit is almost circular. {\bf c} The variation of the distance between $P_1$ and $P_2$ of the orbits of panels (a) and (b) for one revolution.}  
\label{rel-orb}
\end{figure}

The continuous evolution along the family of periodic orbits, from planetary to satellite orbits, is shown in Fig. \ref{orbits123} in three typical orbits on the planetary part ({\it orbit 1}, {\it orbit 2} and {\it orbit 3-0}) and an orbit on the satellite part ({\it orbit 3-1}), just after the transition point. This series of orbits shows the transition from planetary to satellite orbits under a continuous process. In Fig. \ref{orbits34} we present the detail of the {\it orbits 3-0} and {\it 3-1} shown in Fig. \ref{orbits123} and also the satellite {\it orbit 4}. 

In Fig. \ref{rel-orb}a we present the relative motion of the planet $P_1$ around the planet $P_2$, at the transition point ({\it orbits 3-0} and {\it 3-1}) and in Fig. \ref{rel-orb}b the relative motion of $P_1$ around $P_2$ for the satellite orbit ({\it orbit 4)}. Note that in this latter case the relative motion is almost circular. Since the mass of $P_2$ is much larger than the mass of $P_1$, $P_2$ can be considered as the planet and $P_1$ the satellite. A careful inspection of Fig. \ref{orbits34}b reveals that $P_2$ revolves around the star in an almost circular orbit and $P_1$ moves around the star in a ``wave like'' orbit. Note that the {\it orbit 3-0} could also be considered as a satellite motion, although the planet-satellite distance is not small.

In Fig. \ref{rel-orb}c we present the variation of the distance $R_{12}$ between $P_1$ and $P_2$, during one revolution, for the transition {\it orbits 3-0} and {\it 3-1}. Note that this variation is quite small. For comparison, we also show the variation of the distance between $P_1$ and $P_2$ for the satellite {\it orbit 4}, where it is clear that this distance is close to zero. 
\begin{figure}
\begin{center}
$\begin{array}{ccc}
\includegraphics[width=5cm,height=4cm]{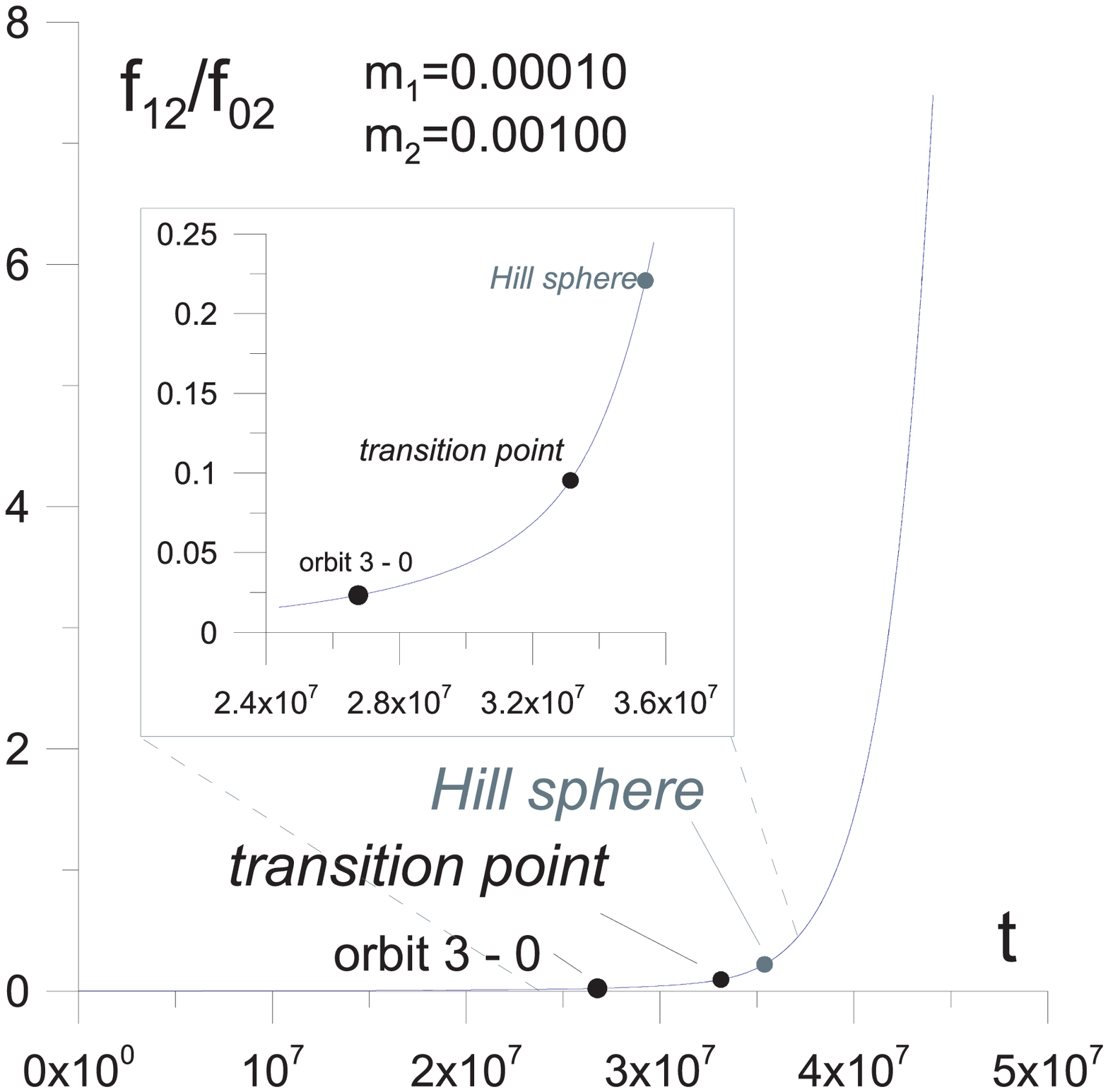} & \qquad&
\includegraphics[width=5cm,height=4cm]{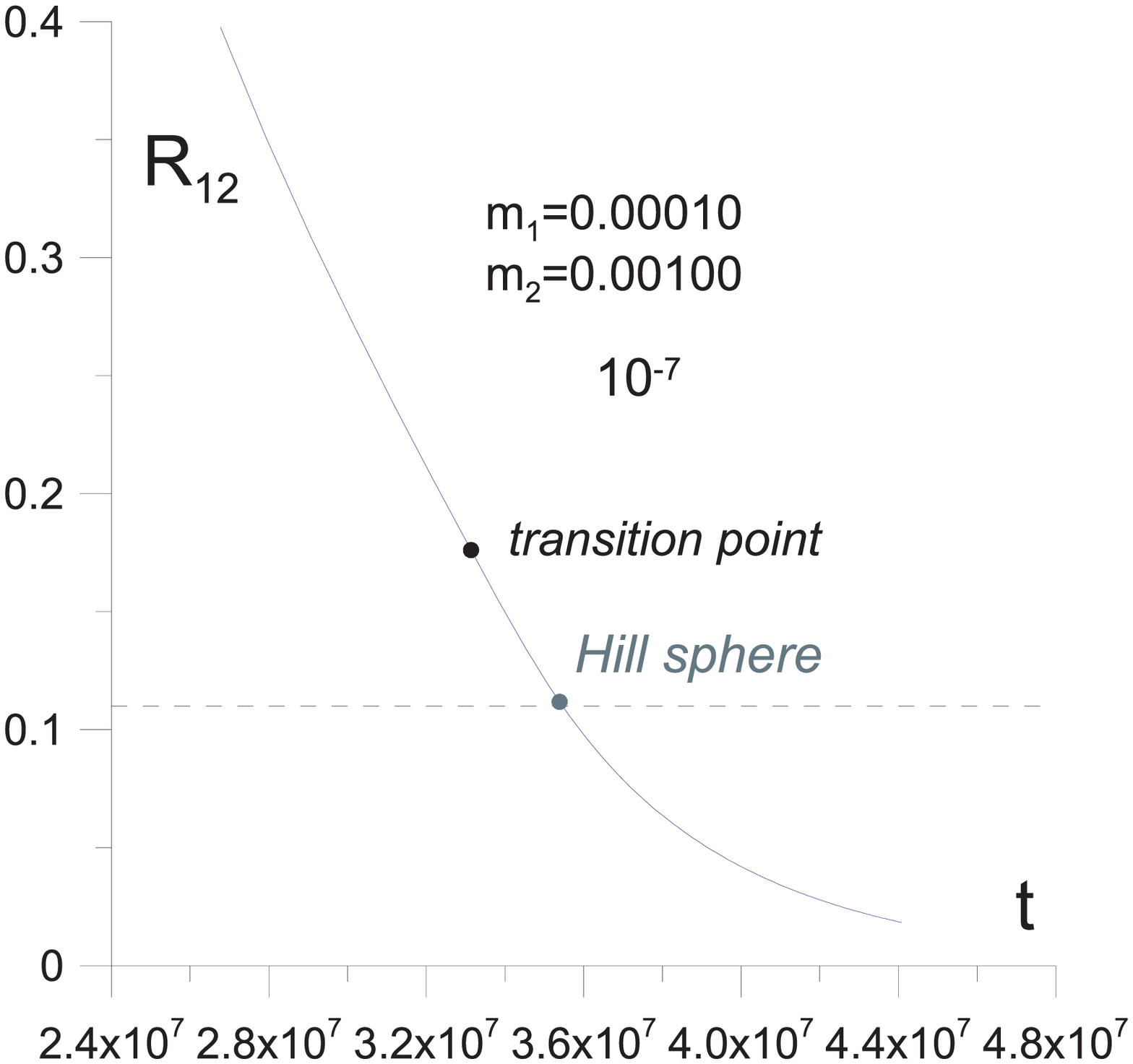}\\
\textnormal{(a)}& \qquad& \textnormal{(b)}
\end{array}$ 
\end{center}
\caption{ {\bf a} The evolution of the ratio $f_{12}/f_{01}$ of the forces between the two small bodies ($f_{12}$) and the attraction from the star ($f_{01}$). The transition point from planetary to satellite orbits is at $t\approx 3.4\times 10^7$. {\bf b} The evolution in time of the distance $R_{12}$ between the two small bodies. The Hill sphere is at the distance $R_{12}=0.11$.}  
\label{Hill}
\end{figure}
\subsection{Transition to satellite orbits and the Hill sphere}

In order to understand the transition from planetary to satellite orbits, along the continuous evolution of the system, we studied the change of the ratio $f_{12}/f_{02}$ of the gravitational forces between the star and $P_2$ ($f_{02})$ and between $P_1$ and $P_2$ ($f_{12}$). This is shown in Fig. \ref{Hill}a. We note that there is a transition point at $t\approx 3.4\times 10^7$, where the ratio $f_{12}/f_{02}$ of the forces increases exponentially, from almost zero values for $t<3.4\times 10^7$. This point coincides with the transition from planetary to satellite orbits, along the evolution of Fig. \ref{e12-dis1}. However this transition is not sharp and an accurate definition of the exact transition point cannot be given in this approach. 

\begin{figure}
\begin{center}
$\begin{array}{ccc}
\includegraphics[width=5cm,height=4.5cm]{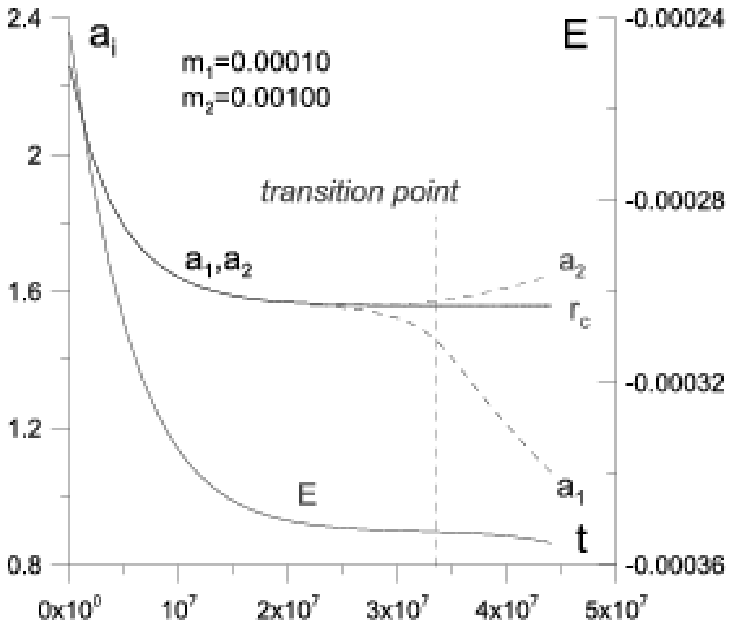} & \qquad&
\includegraphics[width=5cm,height=4.5cm]{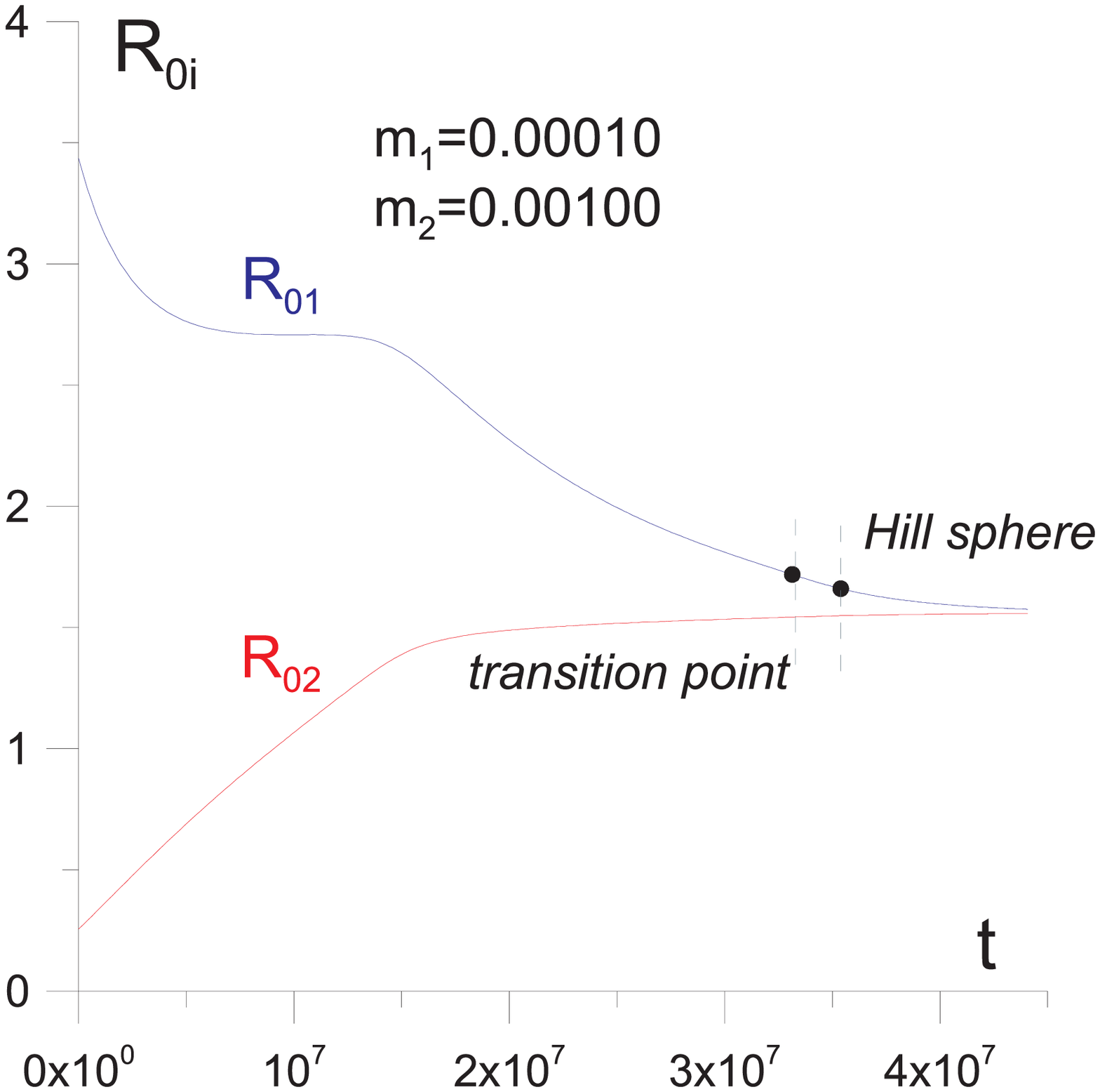}\\
\textnormal{(a)} & \qquad& \textnormal{(b)}
\end{array}$ 
\end{center}
\caption{{\bf a} The change of the semimajor axes $a_1$, $a_2$ and of the total energy $E$ of the system for the evolution from planetary to satellite orbits (see Fig. \ref{e12-dis1}). The radius $r_c\approx 1.56$ of the circular orbit of the center of mass of $P_1$ and $P_2$ is also shown, starting a little before the transition point.  {\bf b} The distances $R_{01}$ and $R_{02}$ of the planets from the star. They coincide at $a_1\approx a_2=1.56$.}  
\label{taE-var}
\end{figure}

The mechanism of transition from planetary motion to trapping in satellite motion is studied by computing the radius, $r_H$, of the Hill sphere along the evolution of the system. This radius, in the circular restricted three body problem approximation, is given by the equation (Murray and Dermott, 1999)
\begin{equation}
r_H=a_2\sqrt[3]{\frac{m_2}{3m_0}},
\label{Hill-1}
\end{equation}
where $a_2$ is the semimajor axis of the planet $P_2$ of mass $m_2$, provided that $P_2$ moves in an almost circular orbit and the mass of $P_1$ is much smaller. We checked that beyond the {\it orbit 3-0} on the planetary part of the evolution (Fig. \ref{e12-dis1}a), the center of mass of $P_1$ and $P_2$ moves around the star in an almost circular orbit with radius $a=1.56$. Also, the motion of the center of mass of the two planets almost coincides with the motion of $P_2$ around the star, as is shown in Fig. \ref{taE-var} and in Figs. \ref{orbits34}a,b, for the {\it orbits 3-0} and {\it 3-1}. The same formula \ref{Hill-1} has been used for the elliptic restricted three body problem, where the semimajor axis $a_2$ is replaced by $a(1-e)$ (Mak\'o et al. 2010). Note however that in the elliptic restricted problem there is no energy integral to restrict the motion.

In the general three body problem model, the formula
\begin{equation}
r_H=\frac{a_1+a_2}{2}\sqrt[3]{\frac{m_1+m_2}{3m_0}},
\label{Hill-2}
\end{equation}
has been used for the radius of the Hill sphere (Smith and Lissauer 2010), where $a_1$, $a_2$ are the semimajor axes of the two planets and $m_1$, $m_2$ their masses. However, in the general three body model there do not exist closed surfaces that restrict the motion (Marchal and Bozis 1982), so the above formula can be considered as an approximation only. In our case we can easily check that both formulas \ref{Hill-1} and \ref{Hill-2} give almost the same results for $r_H$. 

We computed $r_H$ along the evolution of Fig. \ref{e12-dis1}a, starting before the transition point, using Eq. \ref{Hill-1} for $a_2=1.56$, $m_1=0.00010$, $m_2=0.00100$ and $m_0=0.99890$ and we found $r_H=0.11$ (note that after the transition point it is $a_2\approx 1.56$). This is shown in Fig. \ref{Hill}b, where we have marked the position of the transition point ({\it orbit 3}) and the position of the {\it orbit 3-1}, where the distance between $P_1$ and $P_2$ is at the limit of the Hill sphere. From this figure it is clear that beyond the transition point the two bodies $P_1$ and $P_2$ lie inside the Hill sphere, and consequently form a satellite system. We remark that the stability region A shown in Fig. \ref{FigMapxav} starts at a point that corresponds to the cusp at the transition from planetary to satellite orbits. The Hill sphere, which is estimated by the formulas (\ref{Hill-1}) and (\ref{Hill-2}), starts from a point which is located inside the region A. 

In addition to the above, we present in Fig. \ref{taE-var}a the change of the semimajor axes during the migration process. The transition point from planetary to satellite orbits is indicated.  In this figure we also present the evolution of the total energy of the system, which is decreasing. The semimajor axes almost coincide, since we are at the 1/1 resonance, and decrease up to the transition point from planetary to satellite orbits. The curves representing the evolution of $a_1$ and $a_2$ after the transition point (dotted lines) are meaningless because, as we have already mentioned, the gravitational interaction between $P_1$ and $P_2$ dominates the attraction from the star. From this point on we consider the radius $r_c$ of the center of mass of $P_1$, $P_2$ around the star, which is almost constant ($r_c=1.56$).

In Fig. \ref{taE-var}b the distances of $P_1$ ($R_{01}$) and of $P_2$ ($R_{02}$) from the star, along the evolution are presented. They are large along the planetary section, because the eccentricities are large, and also the variation during one revolution is large. After the transition point the planets are trapped in a satellite orbit with their center of mass revolving around the sun in an almost circular orbit with $r_c\approx 1.56$. 

From all the above it is clear that in our case the Hill sphere, computed by \ref{Hill-1} or \ref{Hill-2} is close to the transition area. Thus we come to the conclusion that the trapping of the planetary system to a satellite system along its evolution under the dissipative force,  Eq. \ref{dis-law}, is due to the fact that the two planets eventually come close enough to enter the Hill sphere, and from that point on they form a close binary. However, even before the system reaches the Hill sphere, enters the stability region A where we have ``satellite-like'' motion too.

\begin{figure}
\begin{center}
$\begin{array}{cc}
\includegraphics[width=4cm]{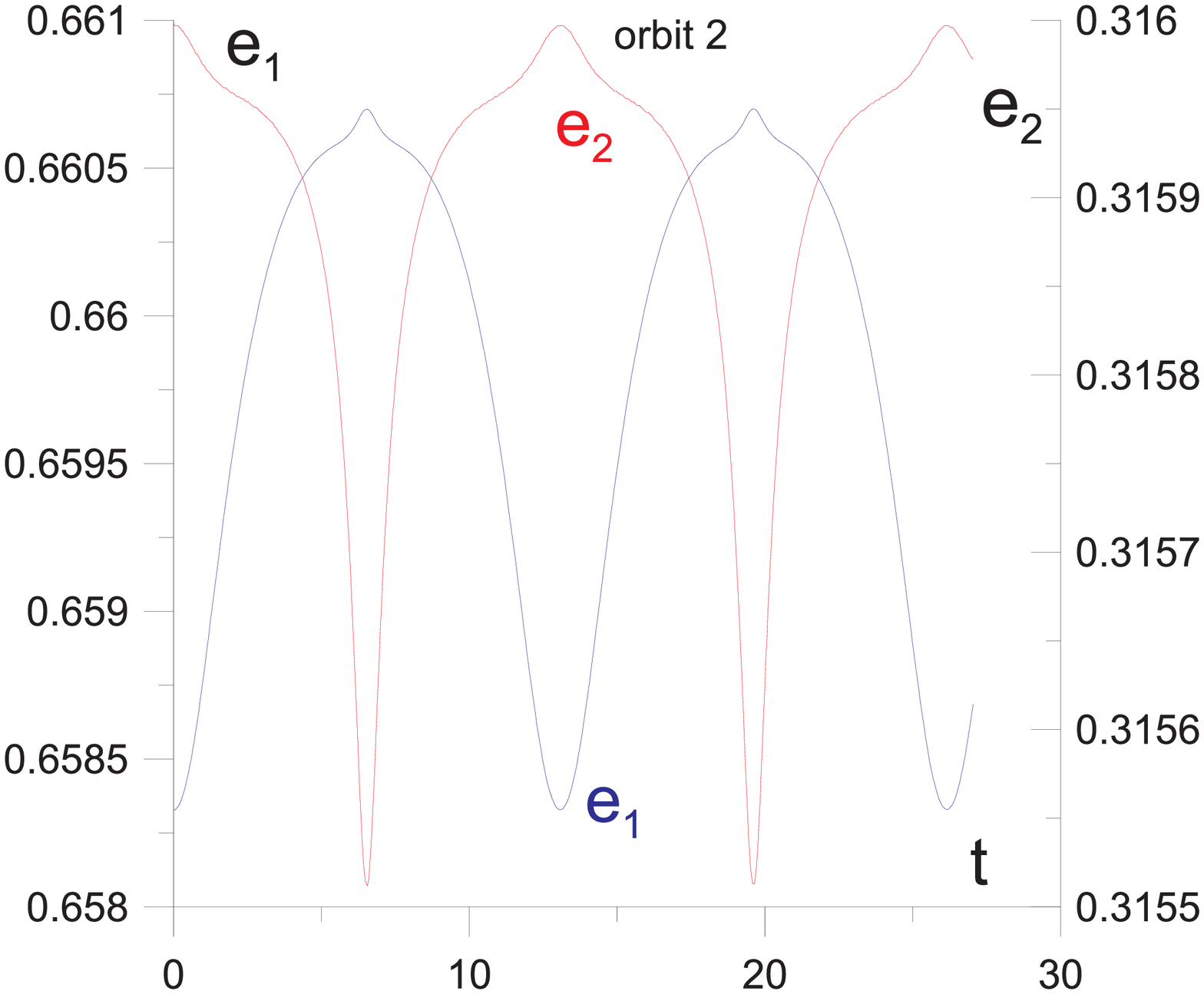} & 
\quad \quad \includegraphics[width=4cm]{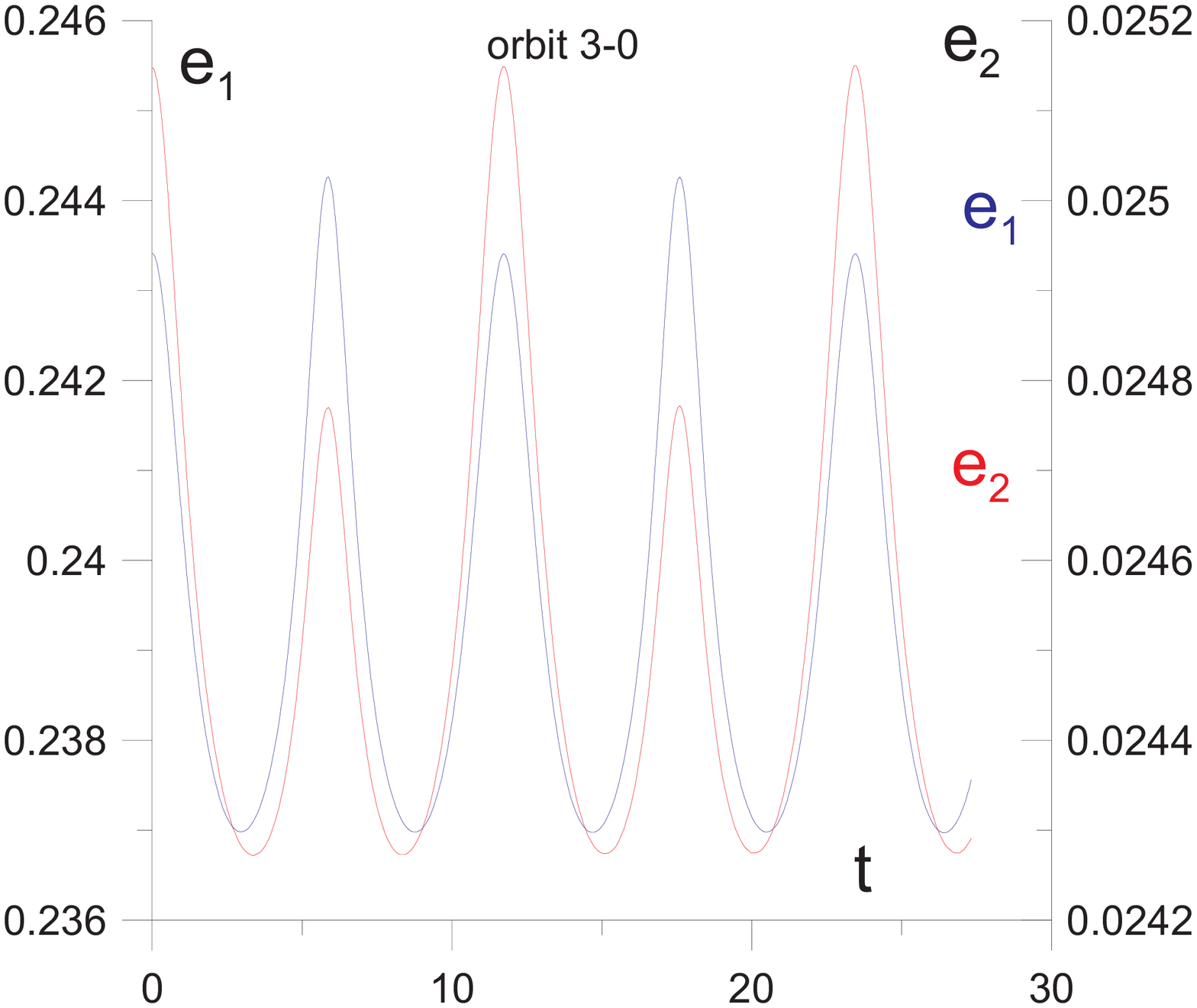}  \\
\textnormal{(a)} & \quad \quad \textnormal{(b)} \\
\includegraphics[width=4cm]{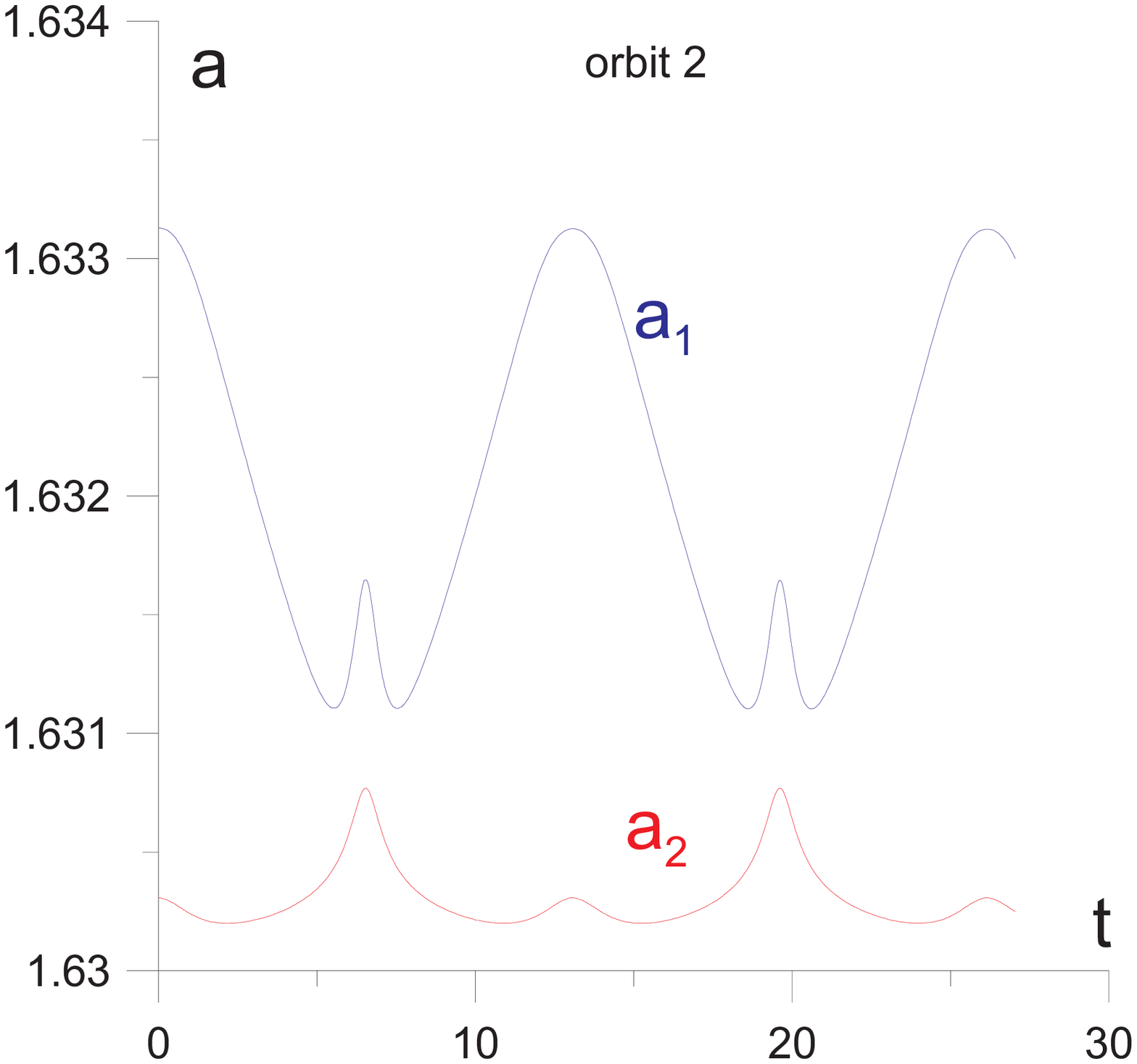} & \quad \quad 
\includegraphics[width=4cm]{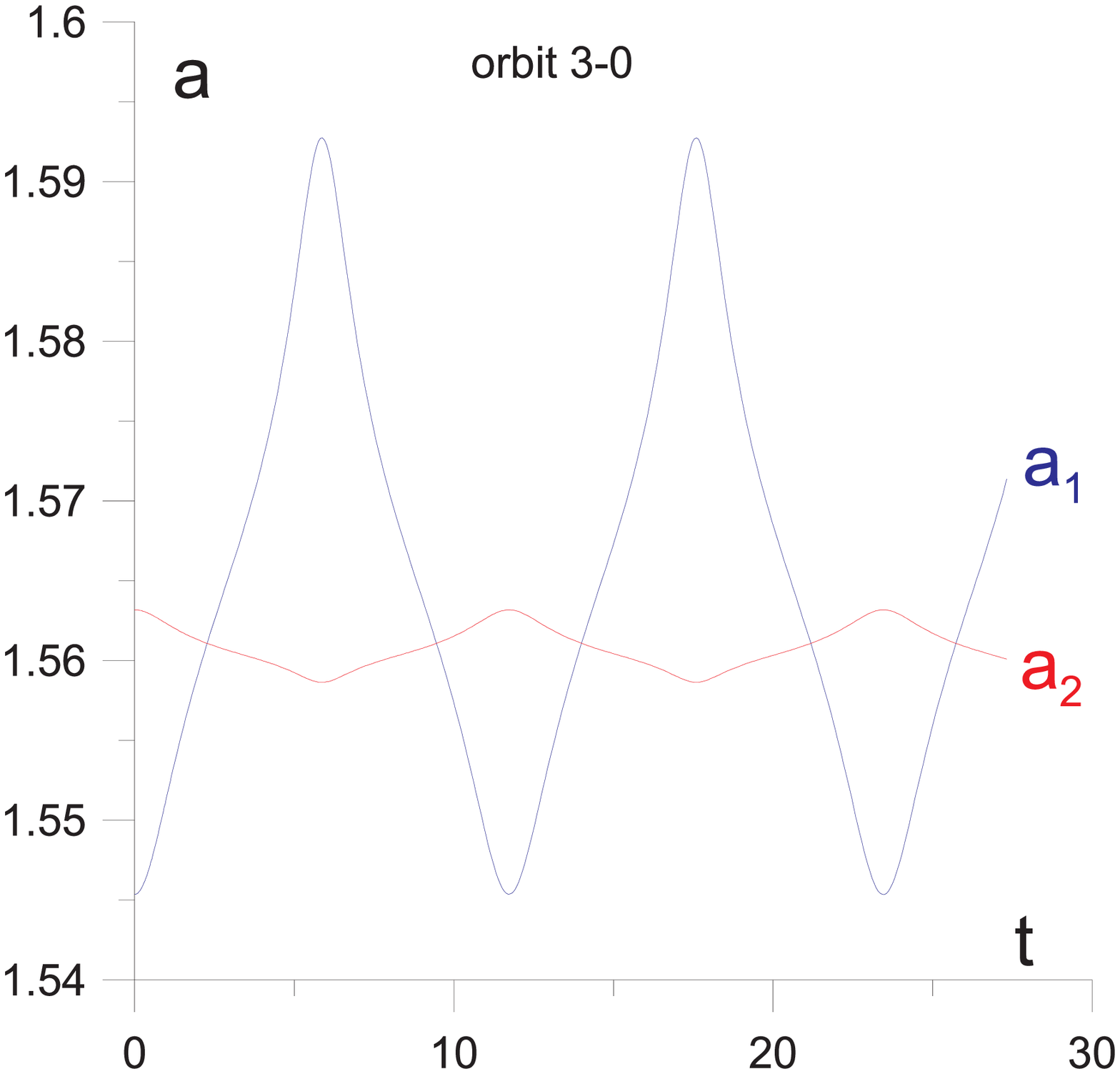} \\ 
\textnormal{(c)}& \quad \quad  \textnormal{(d)}
\end{array}$ 
\end{center}
\caption{{\bf a} The variation of the eccentricities along the planetary {\it orbit 2} of  Fig\ref{e12-dis1}a. {\bf b} The same as in panel (a), for the satellite {\it orbit 3-0} in Fig.\ref{e12-dis1}a. {\bf c} The variation of the semimajor axes of the {\it orbit 2}. {\bf d} The variation of the semimajor axes along the satellite {\it orbit 3-0}.}  
\label{e-var}
\end{figure}

\subsection{Variation of semimajor axes and eccentricities} 
In Fig. \ref{e-var} we present the variation of the semimajor axes and the eccentricities, for one revolution, for two typical orbits along the evolution process shown in Fig. \ref{e12-dis}a. Both orbits are on the planetary part, but {\it orbit 2} has large eccentricities and {\it orbit 3-0} is close to the transition point to satellite motion.  In panels (a) and (c) we present the variation of the eccentricities and the semimajor axes, respectively for the {\it orbit 2} and in panels (b) and (d) the evolution of the eccentricities and the semimajor axes, respectively, for the {\it orbit 3-0}. The behavior is different for these two orbits: For the planetary {\it orbit 2} the maximum of $e_1$ occurs when $e_2$ is minimum, and vice versa. The opposite is true for the {\it orbit 3-0}, where the maximum, or minimum, take place at the same time.

\begin{figure}
\begin{center}
$\begin{array}{ccc}
\includegraphics[width=5cm,height=4.5cm]{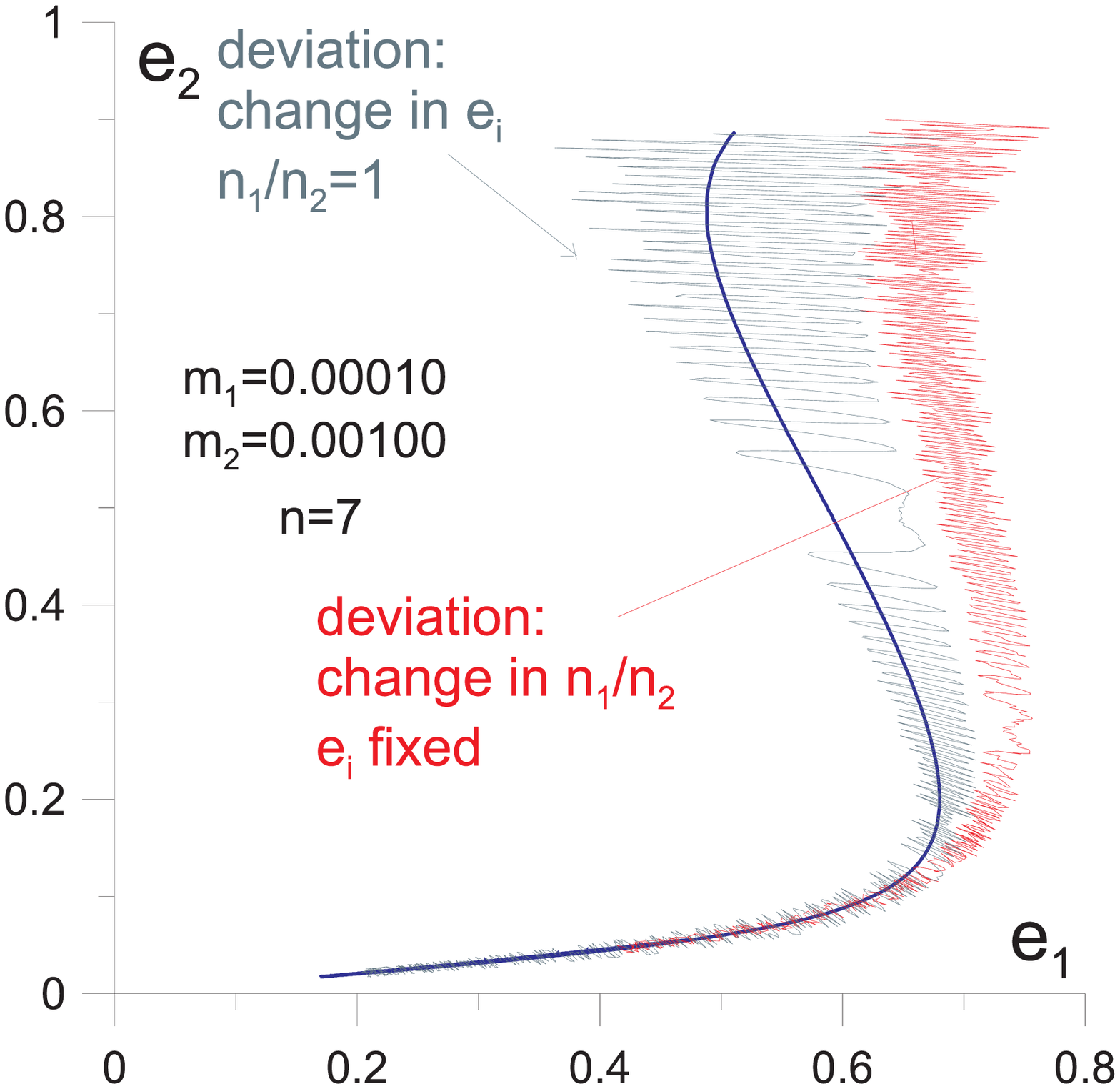} & \qquad&
\includegraphics[width=5cm,height=4.5cm]{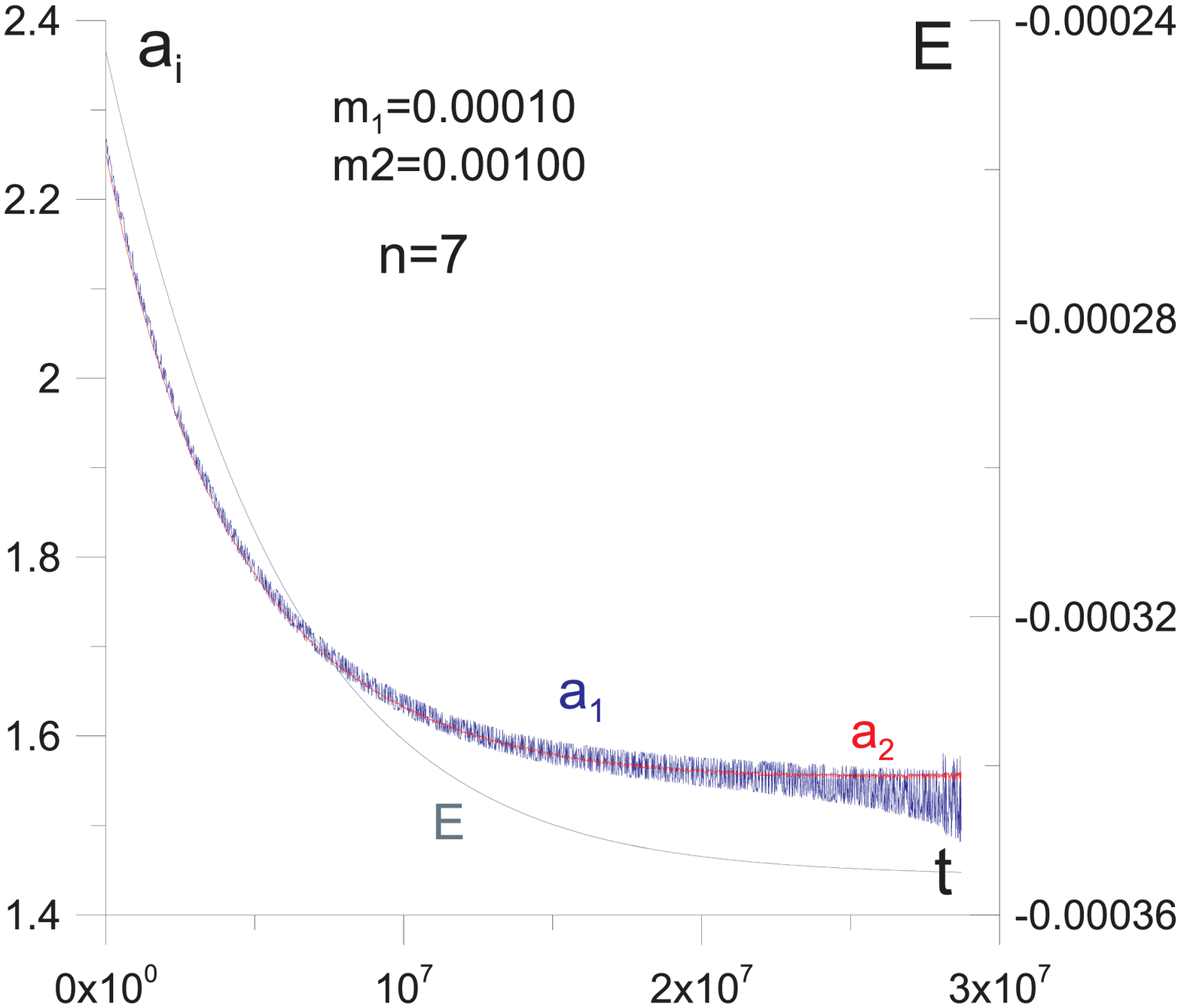}\\
\textnormal{(a)} & \qquad & \textnormal{(b)}
\end{array}$ 
\end{center}
\caption{{\bf a} Typical types of evolution when the starting point is not on the family of periodic orbits. {\bf b} The variation of the planetary semimajor axis and the decreasing of the energy along the evolutions of panel (a).}  
\label{EvolFar}
\end{figure}

\subsection{Starting nearby a periodic orbit}
Another set of numerical studies performed refers to the cases when the starting point is not on the family of periodic orbits, but there is a deviation either in the initial values of the semimajor axes, implying that we are no exactly on the 1/1 resonance (but not far from it), or a deviation in the initial values of the eccentricities, keeping the 1/1 resonance. An example of a typical evolution is shown in Fig. \ref{EvolFar}a.  
We find that the system evolves in this case all the way along the planetary part of the family, librating around the resonant family with decreasing amplitude. Alternatively, the orbit may move at the beginning far from the family, but later is attracted to the family librating around it as before.  However, when the system comes close to the transition point from planetary to satellite orbits, it enters a chaotic region and is finally destabilized. This is due to the fact that the region of stability close to the transition point is very small (see section \ref{Sectionmaps}). In the present case, the smaller body, $P_1$, is ejected from the system and we are left with a stable two-body system consisting of the star and $P_2$. The same evolution appears if there is a deviation in the symmetry, either by changing the angle $\Delta\omega$ between the line of apsides of $P_1$ and $P_2$, up to $\Delta\omega=150^\circ$, or by changing the staring point of $P_2$ at $t=0$ by changing the angle $M_2$, up to $M_2=60^\circ$, from the initial value $M_2=0^\circ$.
In Fig. \ref{EvolFar}b we present the variation of the semimajor axes. The evolution stops close to the transition point, because, as mentioned above, the stability region is very small and the system enters a chaotic zone. 

\section{Conclusions}

We studied the dynamics of planetary and co-orbital motion at the 1/1 resonance and the possibility of the transition from a planetary system, where the two planets revolve around the star in eccentric orbits, to a satellite system where the former two planets form a close pair (planet-satellite) whose center of mass revolves around the star. 

In the conservative general three body problem there exist a family of stable periodic orbits, which are {\it symmetric} with respect to the $x$-axis of the rotating frame $xOy$ given in Fig. \ref{rotating}b. For any planetary mass ratio, the family shows a cusp when it is presented in the eccentricity plane $e_1-e_2$. This cusp separates the family in two parts: one with planetary-like orbits and one with satellite-like orbits. Around any periodic orbit of the above family there is a region of phase space with regular orbits. However at the cusp or, equivalently, at the transition point from planetary to satellite orbits, this region becomes very narrow.  

It is found that the transition from planetary to satellite orbits is possible under a migration process, due to a drag force (Eq. \ref{dis-law}) exerted on the planetary system. The starting point is a resonant periodic orbit, at the 1/1 resonance. This orbit belongs to the family of stable periodic orbits mentioned above.  When the orbit evolves along the {\it planetary part} the gravitational attraction from the star on the two planets dominates the gravitational interaction between these bodies but along the {\it satellite part} the gravitational interaction between the two small bodies dominates. Along the planetary part the eccentricities start with large values and decrease along the family up to the transition point to the satellite part. From that point on the two small bodies form a close binary (a planet-satellite system) whose center of mass revolves around the star in an almost circular orbit. It is important to note that this is a {\it single} family, presented in the space of initial conditions (\ref{initcond}) by a {\it smooth} curve. The system that starts from a periodic orbit of the planetary part evolves, under the drag force, {\it along} the family, all the way and ends to a satellite orbit. This means that it is, in principle, possible to generate a satellite system starting from a planetary system. A typical evolution of this kind is presented in Figs. \ref{e12-dis} and \ref{e12-dis1}.

The above mentioned transition takes place when the starting point of the system is very close to a periodic orbit of the 1/1 resonant stable symmetric family. If the starting point deviates from the exact periodic motion, the evolution of the system follows the planetary part only, up to the transition point (the cusp), but from that point on it follows a chaotic motion, because the region of stability at that part of the resonant family is small. Eventually, the smaller body is ejected and we are left with a two-body system, consisting of the star and the larger of the small bodies. Thus, the transition from the planetary orbits to the satellite orbits, i.e. passage to the region A in phase space (see Fig. \ref{FigMapx1x2} and \ref{FigMapxav}), seems to be of low probability. 

\section{Appendix}
In the following tables we provide the initial conditions of the orbits discussed in section  \ref{NonConEvol} and indicated in figures \ref{e12-dis}a and \ref{e12-dis1}a. 

\begin{table}[ht]
\begin{center}
\caption{Initial conditions in the {\em rotating frame} of the orbits indicated in Fig. \ref{e12-dis}a. For all cases it is $\dot x_1=0$, $y_2=0$, $\dot x_2=0$  ($m_1=m_2=0.001$)}
\label{tf9r}
\begin{tabular}{ccccc}
orbit & $x_1$ & $x_2$ & $\dot y_2$ & $\dot \theta$\\
\hline
1   &  16.714675 & 0.3737924 & 2.2346161 & 0.0032573\\
2   &  7.8752676 & 2.7017427 & 0.6530591 & 0.0323985\\
3   &  5.2407941 & 4.5255771 & 0.1414788 & 0.0783017\\
3-1 &  5.1026149 & 4.6591849 & 0.1160025 & 0.0808149\\
4   &  4.9136557 & 4.8521092 & 0.1849638 & 0.0736204\\
\hline
\end{tabular}
\end{center}
\end{table}

\begin{table}[ht]
\begin{center}
\caption{Initial conditions in the {\em inertial frame} of the orbits given in Table \ref{tf9r}. Initial conditions are given for the bodies $P_1$ and $P_2$. The center of mass of the system (including the star) is in rest at (0,0) and $Y_1=Y_2=0$, $\dot X_1=\dot X_2=0$. }
\label{tf9i}
\begin{tabular}{ccccc}
orbit & $X_1$ & $\dot Y_1$ & $X_2$ & $\dot Y_2$\\
\hline
1   & 16.714302 & 0.0522089 & 0.3734186 & 2.2335978\\
2   & 7.8725659 & 0.2544063 & 2.6990410 & 0.7398509\\  
3   & 5.2362685 & 0.4098672 & 4.5210515 & 0.4953433\\
3-1 & 5.0979557 & 0.4118748 & 4.6545257 & 0.4920415\\
4   & 4.9088036 & 0.3612031 & 4.8472571 & 0.5416358\\
\hline
\end{tabular}
\end{center}
\end{table}

\begin{table}[ht]
\begin{center}
\caption{Initial conditions in the {\em rotating frame} of the orbits indicated in Fig. \ref{e12-dis1}a. For all cases it is $\dot x_1=0$, $y_2=0$, $\dot x_2=0$ ($m_1=0.0001$, $m_2=0.001$) }
\label{tf10r}
\begin{tabular}{ccccc}
orbit & $x_1$ & $x_2$ & $\dot y_2$ & $\dot \theta$\\
\hline
1   & 3.4344156 & 0.2555475 & 2.6871776 & 0.1097631  \\
2   & 2.7079902 & 1.1148626 & 0.9401043 & 0.1310847  \\
3-0 & 1.9213229 & 1.5236791 & 0.3227701 & 0.3263972 \\
3   & 1.7185587 & 1.5425822 & 0.1886959 & 0.4040573 \\
3-1 & 1.6601637 & 1.5483772 & 0.1622589 & 0.4193338 \\
4   & 1.5792598 & 1.5565925 & 0.2294687 & 0.3799935 \\
\hline
\end{tabular}
\end{center}
\end{table}

\begin{table}[ht]
\begin{center}
\caption{Initial conditions in the {\em inertial frame} of the orbits given in Table \ref{tf10i}. Initial conditions are given for the bodies $P_1$ and $P_2$. The center of mass of the system (including the star) is in rest at (0,0) and $Y_1=Y_2=0$, $\dot X_1=\dot X_2=0$. }
\label{tf10i}
\begin{tabular}{ccccc}
orbit & $X_1$ & $\dot Y_1$ & $X_2$ & $\dot Y_2$\\
\hline
1   & 3.4341601 & 0.3742569 & 0.2552920 & 2.7125121\\
2   & 2.7068753 & 0.3538898 & 1.1137477 & 1.0851595\\
3-0 & 1.9197992 & 0.6262943 & 1.5221554 & 0.8192746\\
3   & 1.7170161 & 0.6935842 & 1.5410396 & 0.8111755\\
3-1 & 1.6586153 & 0.6953512 & 1.5468288 & 0.8107342\\
4   & 1.5777032 & 0.5992875 & 1.5550359 & 0.8201428\\
\hline
\end{tabular}
\end{center}
\end{table}

\end{document}